\documentclass[12pt,aps,nofootinbib,preprintnumbers,groupedaddress,letterpaper]{article}

\usepackage{hyperref}
\usepackage[dvipsnames]{xcolor}
\usepackage[normalem]{ulem}
\usepackage{amsmath, bbold}
\usepackage{enumerate}
\usepackage{amsfonts}
\usepackage{yfonts}
\usepackage[textsize=tiny]{todonotes}
\usepackage{cite}

\usepackage{graphicx}

\newcommand\ov{\over }

\def\le{\left}
\def\ri{\right}

\def\({\left(}
\def\){\right)}
\def\<{\langle}
\def\>{\rangle}
\def\Tr{\mathop{\rm Tr}}

\newcommand\half{{\ensuremath{\frac{1}{2}}}}
\newcommand\p{\ensuremath{\partial}}

\newcommand\field[1]{{\ensuremath{\mathbb{{#1}}}}}

\newcommand{\CC}{\field{C}}

\newcommand{\RR}{\field{R}}

\newcommand{\ZZ}{\field{Z}}

\newcommand{\be}{\begin{equation}}
\newcommand{\ee}{\end{equation}}
\newcommand{\bea}{\begin{eqnarray}}
\newcommand{\eea}{\end{eqnarray}}
\newcommand{\bwt}{\begin{widetext}}
\newcommand{\ewt}{\end{widetext}}

\newcommand{\bi}{\begin{itemize}}
\newcommand{\ei}{\end{itemize}}
\newcommand{\ben}{\begin{enumerate}}
\newcommand{\een}{\end{enumerate}}
\newcommand{\bca}{\begin{cases}}
\newcommand{\eca}{\end{cases}}
\newcommand{\bln}{\begin{align}}
\newcommand{\eln}{\end{align}}
\newcommand{\bst}{\begin{split}}
\newcommand{\est}{\end{split}}

\begin{document}

\begin{titlepage}

\begin{flushright}
QMUL-PH-21-41\\
\end{flushright}

\vspace{5mm}

  \begin{center}

\centerline{\Large \bf {Segmented strings, brane tilings,}}
\vskip 0.3cm
\centerline{\Large \bf {and the Y-system}}

\bigskip
\bigskip

{\bf David Vegh}

\bigskip

\small{
{ \it   Centre for Theoretical Physics, Department of Physics and Astronomy \\
Queen Mary University of London, 327 Mile End Road, London E1 4NS, UK}}

\medskip

{\it email:} \texttt{d.vegh@qmul.ac.uk}

\medskip

{\it \today}

\bigskip

%\date{\today}

\begin{abstract}

I show that the motion of a closed string consisting of $n$ segments in AdS$_3$ can be embedded into the mutation dynamics of the  
$Y^{n,0}$ brane tiling.
The determinant of the Kasteleyn matrix computes the spectral curve.
The dynamics is governed by a Y-system with additional constraints ensuring that the string closes in target space. The constraints can be deformed by  coupling the worldsheet to a background two-form whose field strength is proportional to the volume form.

\end{abstract}

\end{center}

\end{titlepage}

\vskip-1.5cm
\tableofcontents

\clearpage

\section{Introduction}

In this paper, we discuss segmented (discretized) strings in three-dimensional anti-de Sitter (AdS) spacetime \cite{Vegh:2015ska, Callebaut:2015fsa, Vegh:2015yua, Vegh:2016hwq, Gubser:2016wno, Gubser:2016zyw, Vegh:2016fcm, Vegh:2018dda, Vegh:2019any, Vegh:2021jga, Vegh:2021jhl}\footnote{For earlier works on discrete strings see \cite{Giles1977, Klebanov:1988ba}.}.  In a flat space limit, these well-known classical solutions describe piecewise linear strings. In AdS$_3$, the worldsheet is built from elementary patches, which in the simplest case are AdS$_2$ subspaces of the target space. The worldsheet conformal field theory (CFT) is integrable, and a suitable exact discretization preserves the integrability structure which we intend to use.

In recent years it has become clear that gravity and quantum physics are not incompatible. In fact, they are intimately connected: the emergence of spacetime is thought to be related to the quantum entanglement of degrees of freedom in a dual quantum system. However, the nature of this emergence is not fully understood and toy models for emergent gravity (such as the Sachdev-Ye-Kitaev model \cite{sy, kitaev, Maldacena:2016hyu}) are valuable.
An interesting toy model is the worldsheet theory of the string itself. In fact, it is considered to be one of the simplest theories of quantum gravity\footnote{The theory of massless free bosons can be perturbed by Zamolodchikov's composite  $T\bar T$  operator\cite{Zamolodchikov:2004ce}, which induces a flow to the (static gauge) Nambu-Goto action of a string in flat space \cite{Cavaglia:2016oda}. The deformation can be regarded as a `gravitational dressing' of the S-matrix.}. Unlike Jackiw–Teitelboim gravity \cite{Teitelboim:1983ux, Jackiw:1984je}, it has local degrees of freedom.
The theory shares interesting similarities with theories of quantum gravity,
e.g. the absence of local off-shell observables, a Hagedorn temperature, a minimal length, and integrable toy black holes \cite{Dubovsky:2012wk}. The worldsheet theory saturates \cite{deBoer:2017xdk} the conjectured bound on chaos \cite{Maldacena:2015waa}. It is tempting to speculate that there exists a (0+1)-dimensional boundary dual to the worldsheet theory of a string hanging from the AdS boundary.

Segmented strings may play a role in the AdS/CFT duality \cite{Maldacena:1997re, Gubser:1998bc, Witten:1998qj}, which is another motivation for this work. The CFT in the planar limit and the bulk sigma model are both integrable \cite{Bena:2003wd, Minahan:2002ve}, which allows for non-trivial tests of the correspondence. Smooth string solutions in AdS$_3$ can be approximated to arbitrary accuracy with segmented strings. During (classical) time evolution the number of segments is a constant and therefore one may restrict attention to such solutions, which presumably constitute an interesting subset of all possible string states.

Euclidean extremal surfaces in AdS are relevant for computing gluon scattering amplitudes in the boundary theory at strong coupling \cite{Alday:2007hr}.
Open strings ending on the boundary are dual to Wilson loops in the boundary theory \cite{Maldacena:1998im, Rey:1998ik}. Segmented strings may be used for generating interesting bulk string solutions in this context.

Spectral curves are important invariants of integrable models. They are they main object in finite-gap constructions, which generalize the inverse scattering method for periodic field configurations. The spectral curve  encodes information about the constants of motion and it describes half of the classical phase space. Quantization of the models can be understood by considering `quantum spectral curves'.
The purpose of the present work is to simplify the calculation of the spectral curve of segmented strings, which has previosuly been computed  by applying the method of matched asymptotic expansions \cite{Vegh:2021jhl}.

The organization of the paper is as follows. In section 2, we construct segmented strings in AdS$_3$. Their motion is described by {\it celestial variables}, which satisfy a simple non-linear equation of motion. We then proceed to explain how to reconstruct the string embedding based on the information stored in the celestial fields. In the last part of the section, we compute the spectral curve in terms of these fields.

Section 3 discusses the cluster integrable theories of Goncharov and Kenyon  \cite{Goncharov2013}. These are associated to bipartite graphs drawn on a torus, a.k.a. {\it brane tilings} \cite{Hanany:2005ve, Franco:2005rj, Hanany:2005ss, Feng:2005gw}. After discussing some tiling technology, we define the (quantum) integrable models and show how to compute their spectral curves.

Section 4 explains how to embed the discrete theory of segmented strings into the cluster transformation dynamics of $Y^{n,0}$ brane tilings. The corresponding Goncharov-Kenyon integrable model is the relativistic Toda model. Its dynamical variables are calculated from the celestial fields of the string. The variables form a {\it Y-system}, which is a ubiquitous structure in the study of integrable models. In the last part of the section, we compute the spectral curve of segmented strings using brane tiling techniques.

Section 5 discusses in detail the simplest case: a string with four segments. The spectral curve is computed and (successfully) matched with previous results in \cite{Vegh:2021jhl}. We also explicitly derive the constraints on the celestial fields which must be satisfied in order for the string to close in AdS$_3$ (without it the string would have an $SO(2,2)$ monodromy as $\sigma \to \sigma + 2\pi$).

Section 6 constructs $SL(2)$-invariant $2\times 2$ Lax matrices and expresses (some of) the constraints in terms of the dynamical variables of the Goncharov-Kenyon integrable model.

Section 7 discusses the effect of coupling the string to a background B-field whose field strength is the volume form. This coupling does not break the continuous symmetries and as a result, the equation of motion remains. However, the constraints do change and we derive the modified formulas.

The paper ends with a discussion section and an appendix containing the perfect matchings of the $Y^{4,0}$ brane tiling, which is relevant for the four-segmented case.

\clearpage

\begin{figure}[h]
\begin{center}
\includegraphics[width=5cm]{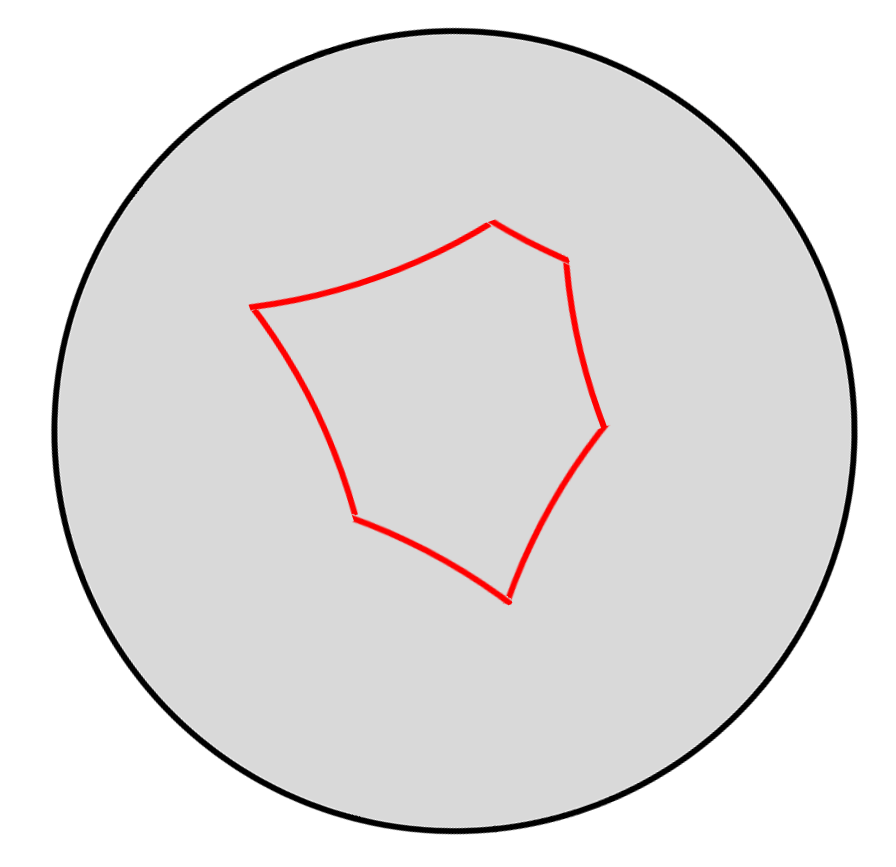} \qquad\qquad
\includegraphics[width=6cm]{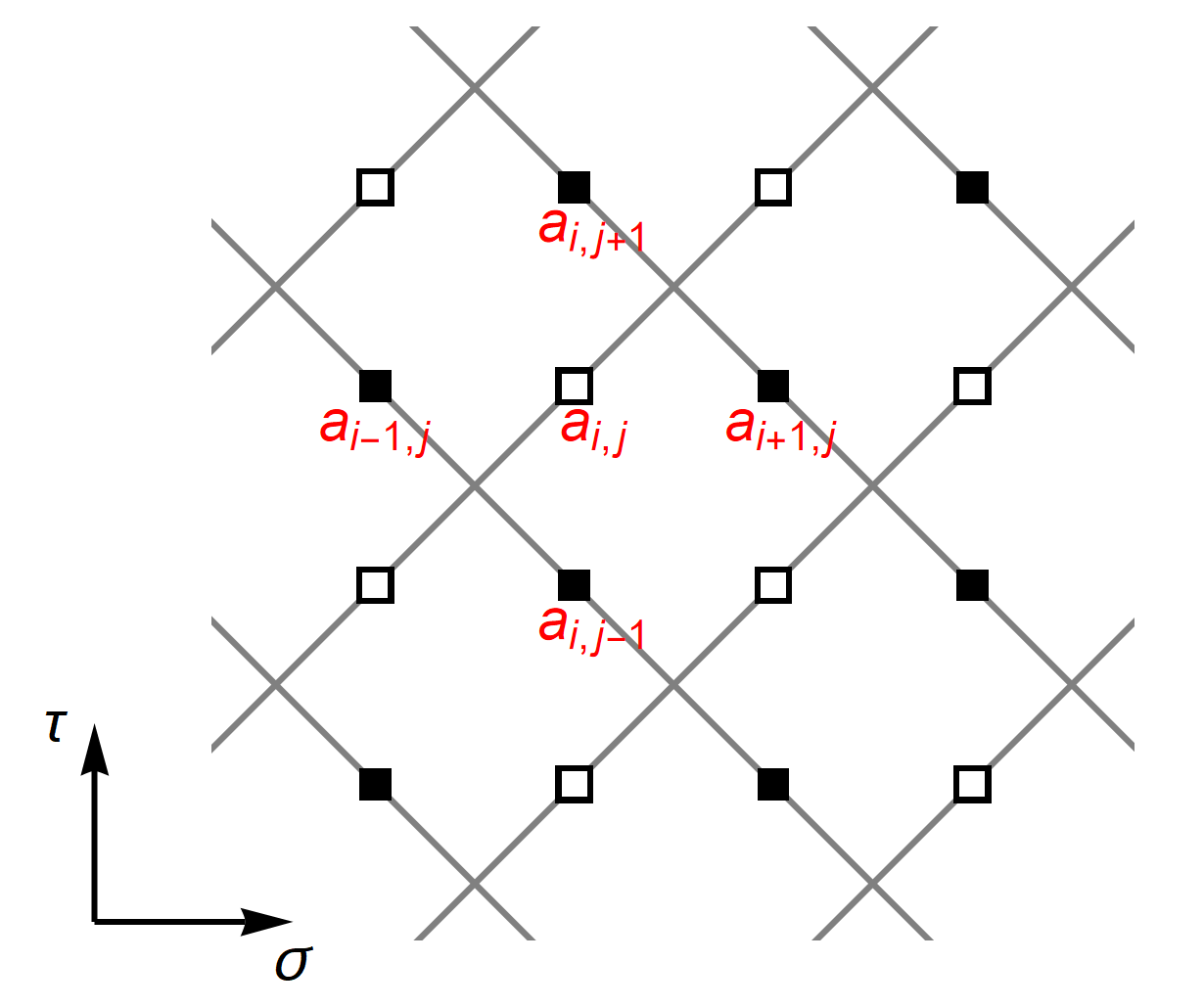}
\caption{\label{fig:sub} {\it Left:} A string consisting of six segments on a timeslice of global AdS$_3$. Kinks between adjacent segments move with the speed of light. {\it Right:}  The string worldsheet is parametrized by the usual $\tau, \sigma$ coordinates. Kink worldlines (lines drawn at 45$^\circ$ angles) form a rectangular lattice on the worldsheet. Between two collisions, the velocity of a kink is characterized by a constant null tangent vector $X \in \RR^{2,2}$. The corresponding celestial field $a= {X_{-1} + X_2 \over X_0 + X_1}$ can be written on the edges, indicated by black and white squares depending on edge orientation.
}
\end{center}
\end{figure}

\section{Segmented strings in AdS$_3$}

In this section we discuss segmented Nambu-Goto strings in AdS$_3$.
A unit size AdS$_3$ can be immersed into the $\RR^{2,2}$ ambient space via the equation
\be
 % \nonumber
   \label{eq:hyp}
   Y \cdot  Y \equiv -Y_{-1}^2 - Y_0^2 + Y_1^2 + Y_2^2 = -1 \ , \qquad  Y \in \RR^{2,2}.
\ee 
We write the sigma model action as
\be
   \nonumber
 S  = -{T \ov 2}\int d\tau d\sigma ( \p_\sigma Y^\mu \p_\sigma Y_\mu - \p_\tau Y^\mu \p_\tau Y_\mu + \lambda( Y^2 + 1)) ,
\ee
where the $ Y(\tau, \sigma)$   function maps the worldsheet into target space.  $T$ is the string tension  and $\lambda$ is a Lagrange multiplier that is used to map the string onto the hyperboloid (\ref{eq:hyp}).
In lightcone coordinates
\be
  \nonumber
  z = \half(\tau-\sigma) \ , \quad \bar z = \half(\tau+\sigma)  \ , \quad  \p \equiv \p_z = \p_\tau - \p_\sigma  \ , \quad \bar\p \equiv \p_{\bar z} = \p_\tau + \p_\sigma
\ee
the equation of motion takes on a simple form
\be
   \nonumber
 \label{eq:eoms}
  \p \bar\p Y - (\p Y \cdot \bar\p Y )  Y = 0 \, .
\ee
The equations are supplemented by the Virasoro constraints
\be
  \label{eq:virasoro}
  \p Y \cdot \p Y = \bar\p Y \cdot \bar\p Y = 0 \, .
\ee
These equations have many (almost everywhere) smooth solutions. Henceforth we restrict our attention to {\it segmented strings} which form a dense set in the space of all string solutions\footnote{FIG. 2 in \cite{Vegh:2021jga} explains how to approximate a smooth solution to arbitrary accuracy.}. Segmented strings are AdS generalizations of piecewise linear strings in Minkowski spacetime.  Each string segment is a linear subspace in $\RR^{2,2}$ \cite{Vegh:2015ska, Callebaut:2015fsa}. Figure \ref{fig:sub} (left) shows the example of a closed string with six segments.
Kink worldlines form a quad lattice on the worldsheet as seen in Figure \ref{fig:sub} (right).
Each diamond in the figure corresponds to a patch of AdS$_2$ characterized by a constant normal vector. Time evolution of the normal vectors (or the kink collision vertices) can be computed by applying a reflection formula \cite{Vegh:2015ska, Callebaut:2015fsa, Vegh:2016hwq, Gubser:2016wno, Vegh:2016fcm}.

\subsection{Celestial variables}
\label{sec:cele}

Segmented strings can equivalently be described by a discrete-time  Toda lattice \cite{Vegh:2016hwq}. Using an $\RR^{2,2}$  version of the spinor-helicity formalism, the kink velocity vectors can be decomposed into products of spinor pairs. The area of the worldsheet can be expressed in terms of these spinors which then leads to the equation of motion
\be
  \label{eq:deqn}
  \hskip -0.15cm {1\ov a_{ij} - a_{i,j+1}}+   {1\ov a_{ij} - a_{i,j-1}} =
  {1\ov a_{ij} - a_{i+1,j}}+   {1\ov a_{ij} - a_{i-1,j}} \, .
\ee
Here $i, j$ are integer indices labeling kink worldlines as shown in Figure \ref{fig:sub} (right). In the case of a closed string, the $i$ index is cyclic. Black and white squares indicate  left- and right-moving kinks. The discrete field $a_{ij}\in \RR$ is the left-handed {\it celestial variable}. Left- and right-handed celestial variables are computed by the formulas
\be
  \label{eq:aeq}
  {a} = {X_{-1} + X_2 \over X_0 + X_1}  \, , \qquad\qquad
  {\tilde a} = {X_{-1} + X_2 \over -X_0 + X_1}  \, ,
\ee
where $X$ is the null tangent vector of the kink. Both fields satisfy (\ref{eq:deqn}). In the following, we will focus on left-handed celestial fields $a_{ij}$ since they contain sufficient information for the reconstruction of the string embedding (up to an $SL(2)$ transformation).

 Initial conditions for (\ref{eq:deqn}) can be given by specifying two rows of celestial variables, e.g. $a_{i,0}$ and $a_{i,1}$ for all $i$. Note that for a Lorentzian embedding, the celestial variables satisfy inequality constraints ensuring that the patch areas are non-negative.

\begin{figure}[h]
\begin{center}
\includegraphics[width=4cm]{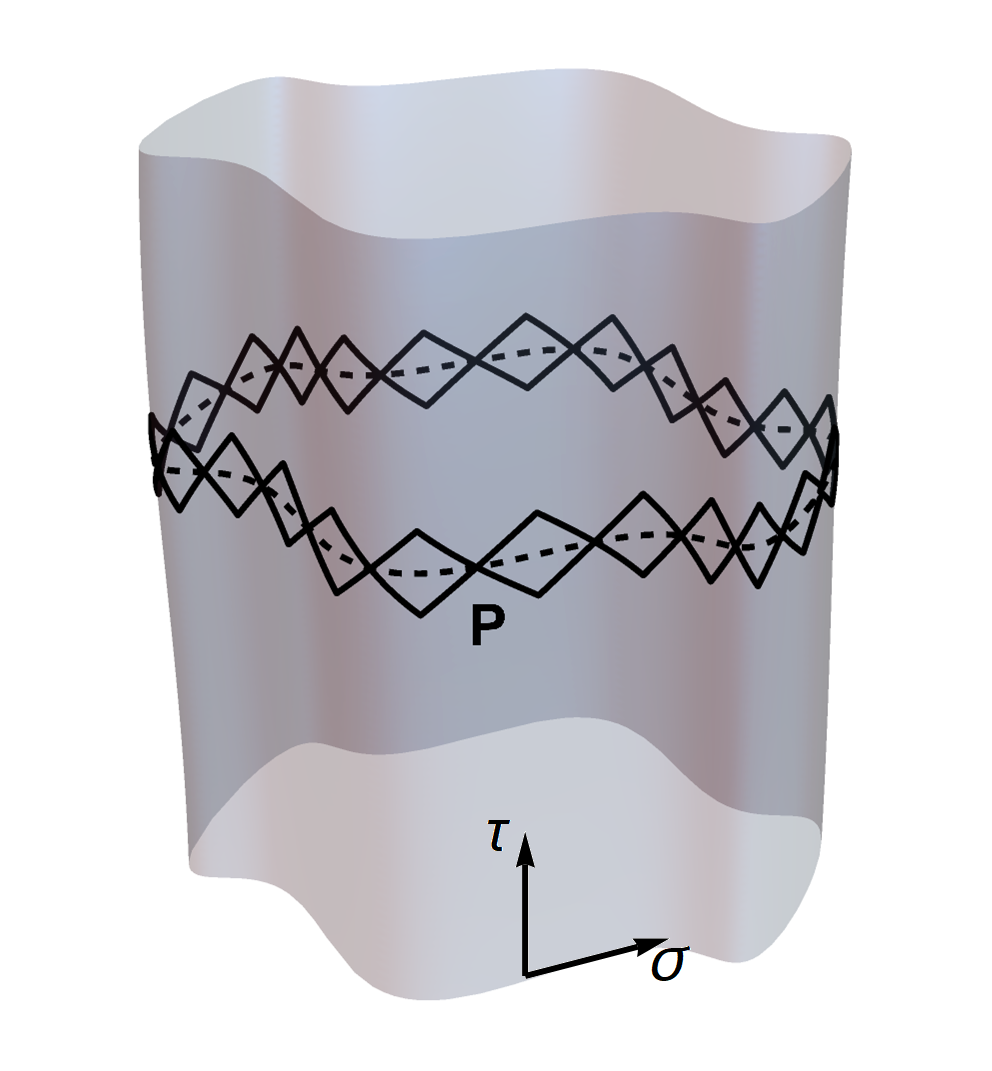} \qquad
\includegraphics[width=11cm]{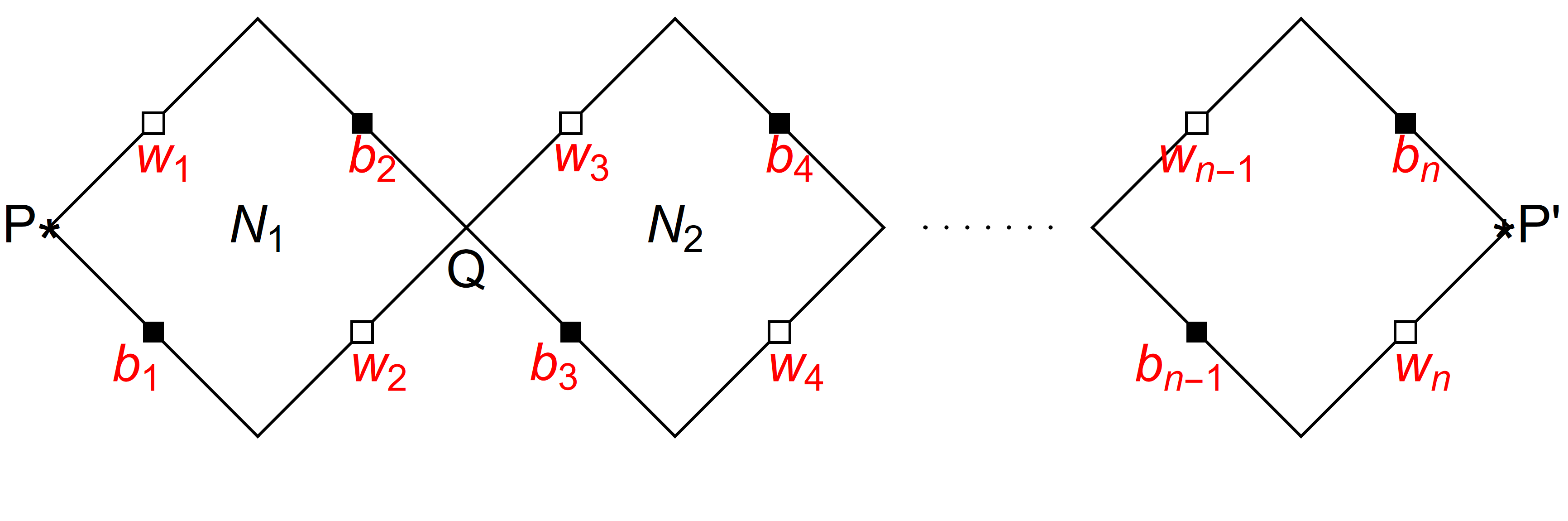}
\caption{\label{fig:spectral}{\it Left:} Schematic drawing of the worldsheet of a closed string in target space. Kink worldlines are indicated near a timeslice (dashed loop).  {\it Right:} The worldsheet of a segmented string consisting of $n$~segments. $P, Q, P' \in \RR^{2,2}$ denote kink collision vertices (we assume $P = P'$ for a closed string). $N_i$ denote AdS$_2$ normal vectors. $b_{i},w_{j} $ are celestial variables.
}
\end{center}
\end{figure}

\subsection{Reconstructing the embedding}

\label{sec:recons}

The string embedding can be obtained from the celestial variables up to a global transformation as follows \cite{Vegh:2016hwq}.
Let us define the `reflection matrix' \cite{Vegh:2019any}
\be
\label{eq:reflmat}
\mathcal{R}_{b,w} =
{1\ov b-w}
\begin{pmatrix}
0 & bw+1 & bw-1 & -b-w
\cr
-1-bw & 0 & b+w & bw-1
\cr
-1+bw & b+w & 0 & -1-bw
\cr
-b-w & bw-1 & bw+1 & 0
\end{pmatrix}   \, .
\ee
Let us assume that all celestial variables are known, along with the position vector of a single vertex  $P \in \RR^{2,2}$ as shown in Figure \ref{fig:spectral}. The   normal vector $N_1$ of the adjacent AdS$_2$ patch can be computed via
\be
   \label{eq:nemferde}
    N_1 =  \mathcal{R}_{b_1,w_1} P \, .
\ee
From the normal vector and the celestial variables, all four vertices can be computed. For instance,
\be
   \nonumber
    Q =  \mathcal{R}_{b_2,w_2} N_1 \, .
\ee
By continuing this procedure, all vertices (and normal vectors) can be determined which then fixes the string embedding in AdS$_3$.

Note that on a closed string the $b_i, w_j$ celestial variables are not completely arbitrary because the string must close in target space. This is ensured if the celestial variables satisfy
 three real constraint equations as we will see.

\subsection{The spectral curve}
\label{sec:spectral}

The discrete Lax matrix for a path that starts on a kink collision vertex and ends in the interior of the adjacent AdS$_2$ patch is given by \cite{Vegh:2021jhl}
\be
  \label{eq:mono}
\Omega_{b,w}(\zeta) =
{1 \over (b-w) \zeta}
\begin{pmatrix}
{b \zeta^2 - w } & {b w(1-\zeta^2)  }
\cr
 { \zeta^2-1  } & { b -w\zeta^2  }
\end{pmatrix}
\ee
Here $\zeta$ is the spectral parameter and $b, w$ are the two adjacent celestial variables. For instance in Figure  \ref{fig:spectral} (right) the connection between $P$ and $N_1$ is given by $\Omega_{b_1,w_1}(\zeta)$.

The Lax matrix satisfies
\be
  \nonumber
   \det \Omega_{b,w} = 1 \, , \qquad \Omega_{b,w}(\zeta=1) = \mathbb{1} \, , \qquad
    \Omega_{b,w}^{-1}(\zeta) =   \Omega_{w,b}(\zeta)  =   \Omega_{b,w}(\zeta^{-1})\, .
 \ee
Simultaneous M\" obius transformations of the celestial variables
\be
  \nonumber
  b \to b' = {a_{00} b + a_{01} \over a_{10} b + a_{11} } \, , \qquad  w \to w' = {a_{00} w + a_{01}  \over a_{10} w + a_{11} }
\ee
are isometry transformations of AdS$_3$. The connection matrix transforms covariantly
\be
  \label{eq:noninv}
  \Omega_{b,w} \to  U \Omega_{b',w'} U^{-1} \, \qquad \textrm{with} \qquad U =  \begin{pmatrix}
a_{00} & a_{01}
\cr
a_{10} & a_{11}
\end{pmatrix}   \, .
\ee
The connection corresponding to longer paths can be computed by multiplying $\Omega$ matrices.
The spectral curve of a closed string can obtained from the monodromy around the non-trivial loop on the string worldsheet. This is shown in Figure \ref{fig:spectral}. Let us assume that  there are an even number of string segments with ${n/ 2}$ left-moving and ${n/ 2}$ right-moving kinks. The monodromy is given by the product of Lax matrices
\be
  \label{eq:monodromy}
\Omega(\zeta) = \Omega_{b_n, w_n}^{-1} \Omega_{b_{n-1}, w_{n-1}} \cdots \Omega_{b_2, w_2}^{-1} \Omega_{b_1, w_1} \, .
\ee
Note that the monodromy depends on where we started  on the worldsheet. In this example the starting (and ending) point was $P$. Changing the starting point changes the monodromy by conjugation.

It has been shown  \cite{Vegh:2021jhl} that  the string closes precisely if
\be
  \label{eq:closeo}
  \Omega(\zeta=i) =   \mathbb{1} \, .
\ee
In general, this criterion is equivalent to imposing three real constraint equations on the
celestial variables.

Finally, the {\it spectral curve} is defined by the equation
\be
 \nonumber
 \det(\lambda I - \Omega(\zeta) ) = 0
\ee
in the space parametrized by $\{ \lambda, \zeta \} \in \CC^2$.
One of the goals of this paper is to compute the spectral curve of segmented strings using an alternative technique based on brane tilings, which will be the topic of the next section.

\clearpage

\begin{figure}[h]
\begin{center}
\includegraphics[width=6cm]{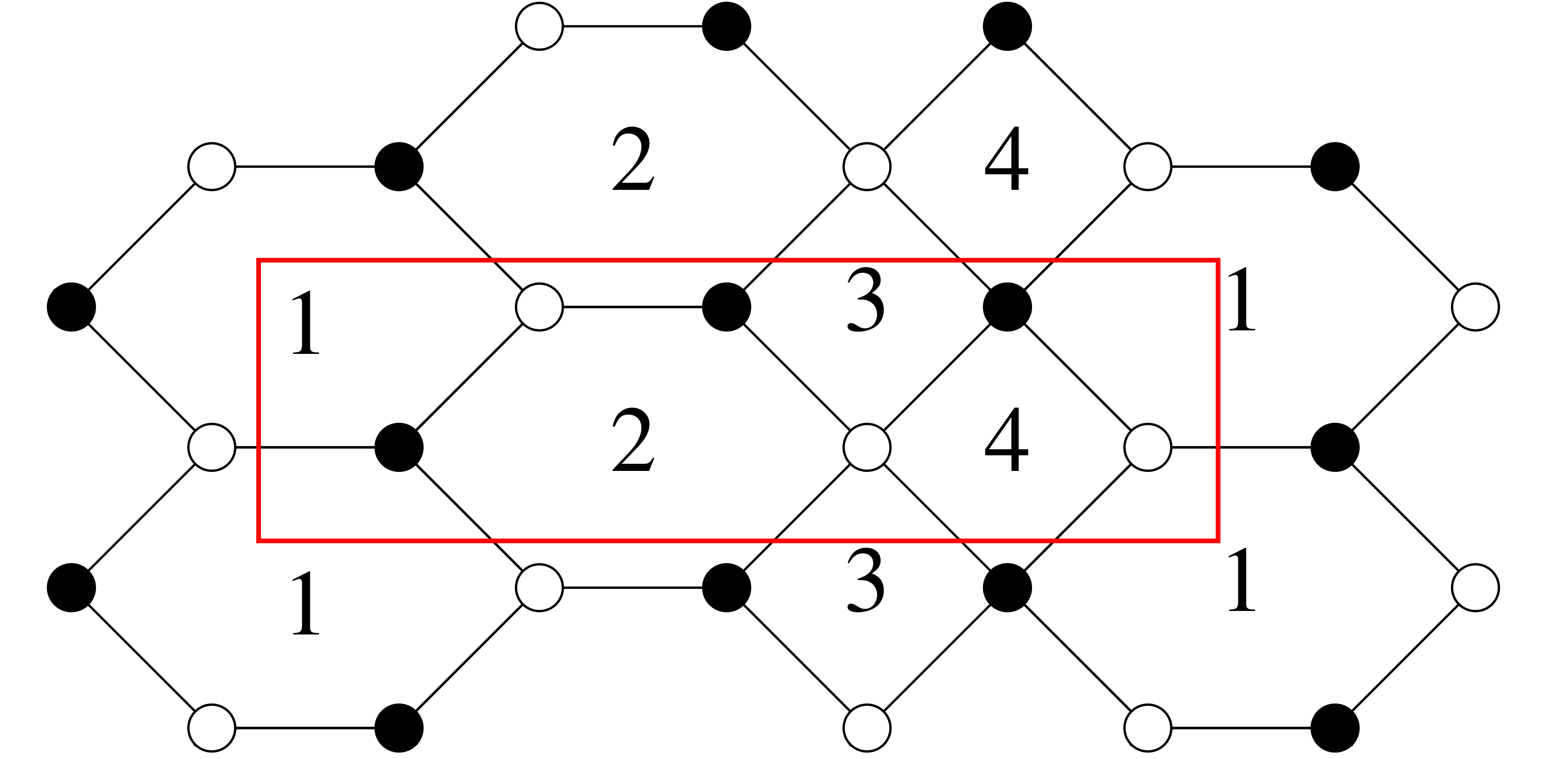} \qquad\qquad
\includegraphics[width=3cm]{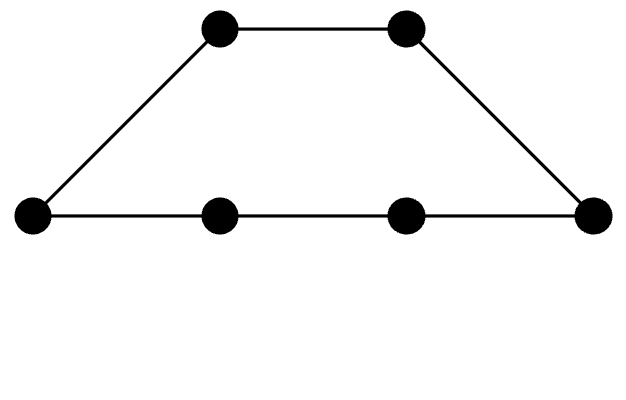}
\caption{\label{fig:l131} {\it Left:} An example brane tiling corresponding to the $L^{131}$ geometry \cite{Cvetic:2005ft, Franco:2005sm}. {\it Right:} The Newton polygon of the dimer model partition function gives the toric diagram of $L^{131}$. This can be straightforwardly computed from the adjacency matrix of the tiling.
}
\end{center}
\end{figure}

\section{Cluster integrable systems}

Goncharov and Kenyon showed \cite{Goncharov2013} that the dimer model on a bipartite graph on a torus gives rise to a quantum integrable system. These special theories have been termed {\it  cluster integrable
systems}. Such bipartite graphs are also called {\it brane tilings} and they previously appeared in the study of four-dimensional quiver gauge theories that arise on the worldvolume of D3-branes probing non-compact toric Calabi-Yau threefolds \cite{Hanany:2005ve, Franco:2005rj, Hanany:2005ss, Feng:2005gw}. Similar graphs have been interpreted as on-shell diagrams in the study of scattering amplitudes  in planar four-dimensional theories \cite{Arkani-Hamed:2012zlh}. In this section we review the construction of cluster integrable models using tilings.

\subsection{Brane tilings}

In this section we give a short introduction to brane tilings \cite{Franco:2005rj}. A planar graph is bipartite if the vertices can be colored black and white, such that edges only connect nodes of different color. A brane tiling is then a doubly periodic bipartite graph\footnote{In the context of 4d $\mathcal{N}=1$ quiver theories, the tiling is the dual graph of the quiver diagram: its faces correspond to gauge groups, edges are chiral bifundamental fields, and the black and white vertices correspond to terms in the superpotential. Tilings are a
generalization of earlier brane box \cite{Hanany:1997tb, Hanany:1998it} and brane diamond models
\cite{Aganagic:1999fe}. }. An example is shown in Figure \ref{fig:l131} (left). A {\it fundamental cell} for the doubly periodic tiling is shown in red. There are four faces and six vertices in this tiling which correspond to the so-called $L^{131}$ geometry.

\begin{figure}[h]
\begin{center}
\includegraphics[width=8cm]{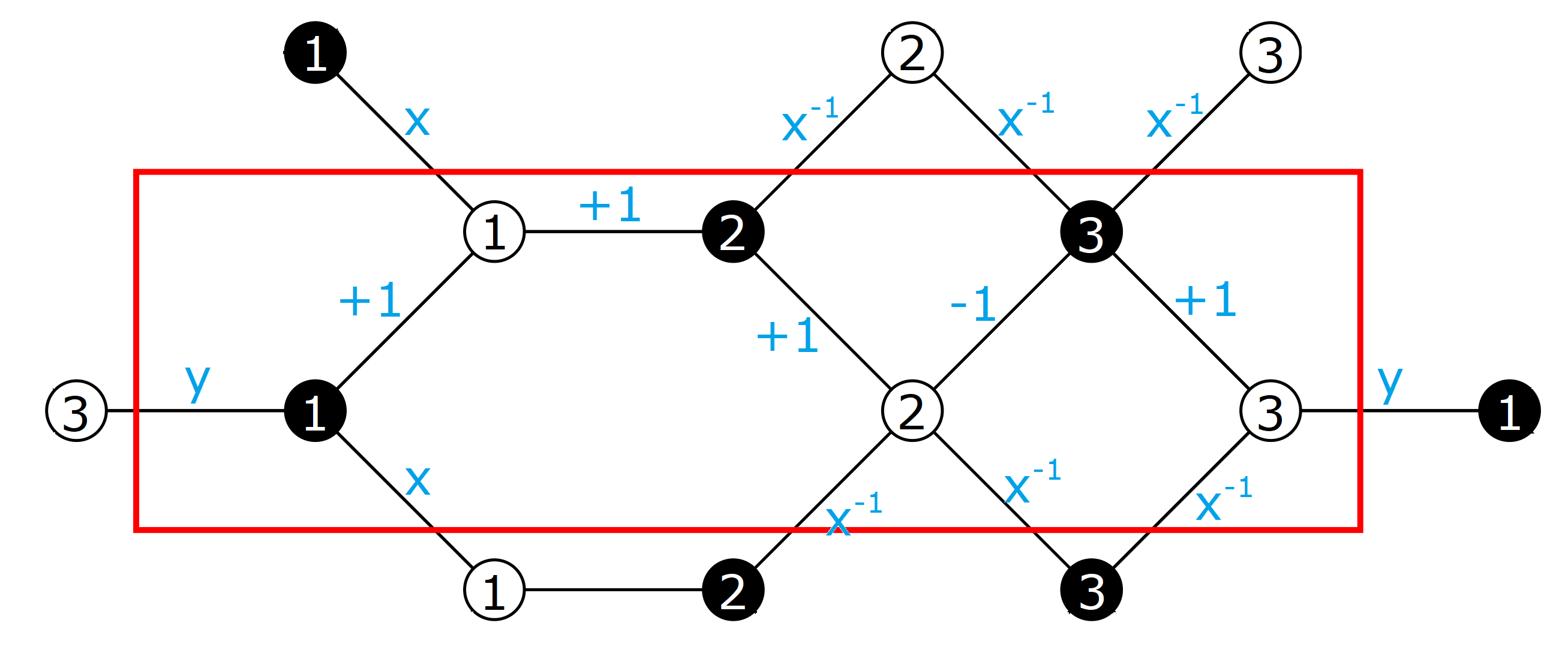}
\caption{\label{fig:l131w} $L^{131}$ brane tiling dressed with edge weights. The three white and three black vertices label the rows and columns of the Kasteleyn matrix, respectively.
}
\end{center}
\end{figure}

A {\it perfect matching} is a subgraph of the tiling that contains all the nodes and each node has
valence one, i.e. edges (dimers) do not touch each other. A set of perfect matchings is given in the appendix. One can assign to the tiling the Newton polygon of the dimer partition function. For the previous example this is shown in Figure \ref{fig:l131} (right). This is a convex lattice polygon which can be computed from the determinant of the {\it Kasteleyn matrix} \cite{Franco:2005rj, Hanany:2005ve, Kenyon:2002a, Kenyon:2003uj}, which is the  adjacency matrix of the tiling graph.  More precisely,
the rows are labeled by the white nodes and the columns by the black nodes. The corresponding entry in the matrix is zero if the two nodes are not connected; otherwise it is the $w_i$ weight of the connecting $i$ edge. If the pair of nodes is connected by more than one edge, then their contributions have to be summed.

The weights are determined as follows. Let us assign a $\pm 1$ weight to every edge in the tiling such that when we multiply the weights for the edges around a face $F$, then we get \bea
\prod_{i \in \{\textrm{edges around } F\}} w_i  =\left\{\begin{array}{l}
+1 \quad \textrm{ if } \quad  e_F=2 \mod 4 \\
-1 \quad \textrm{ if } \quad  e_F=0 \mod 4
 \end{array}\right. \, ,
\eea
where $e_F$ is the number of edges around $F$.
Such an assignment is always possible. Next, we take two non-trivial loops $C_x$ and $C_y$ in the tiling, winding respectively around the $(1,0)$ and $(0,1)$ cycles of the torus. The weight of an edge that is crossed by $C_x$ is multiplied by $x$ or $x^{-1}$ depending on the orientation of the edge (i.e. on which side of $C_x$ does the black vertex lie). Similarly, weights of edges which cross $C_y$ are multiplied by $y$ or $y^{-1}$.

The determinant of the Kasteleyn matrix $P(x,y)=\det K(x,y)$ is a Laurent polynomial. It  is the partition function of perfect matchings of the brane tiling \cite{Kenyon:2003uj}.

In the case of our example, the $L^{131}$  tiling we have the Kasteleyn matrix
\bea
  K=\left( \begin{array}{ccc}
  1+x & 1 & \ 0 \\
  0 & 1+x^{-1} & -1+x^{-1}        \\
  y & 0 & 1+x^{-1}
  \end{array}\right)
\eea
Then
\be
  P(x,y)= 3 + x^{-2} + 3 x^{-1}+x - y + x^{-1} y
\ee

\begin{figure}[h]
\begin{center}
\includegraphics[width=6cm]{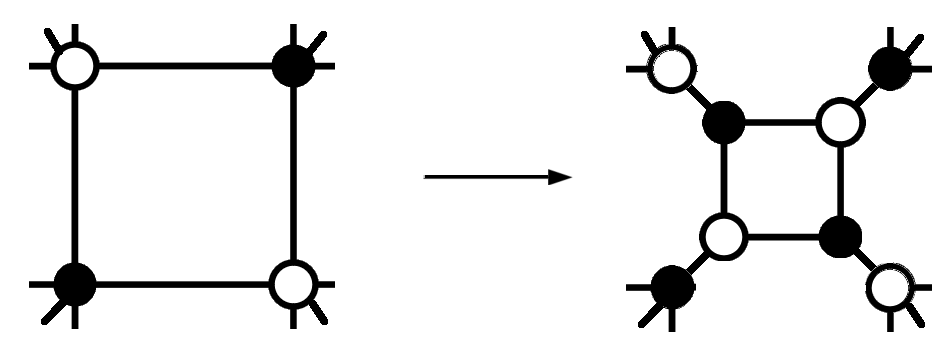} \qquad\qquad\qquad
\includegraphics[width=3.5cm]{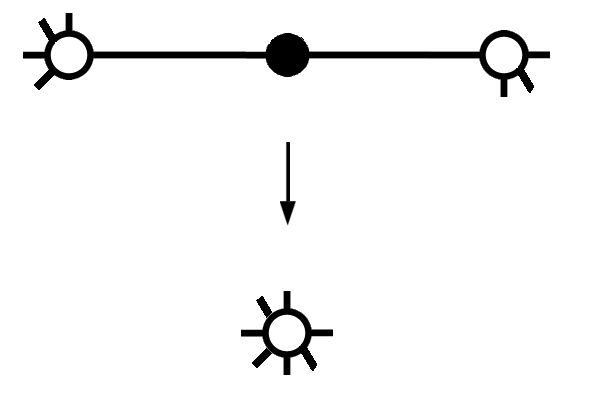}
\caption{\label{fig:seiberg}{\it Left:} `Urban renewal' transformation on the tiling.  {\it Right:} Elimination of valence two nodes. Both transformation preserve the corresponding Newton polygon.
}
\end{center}
\end{figure}

\noindent
whose Newton polygon is shown in Figure \ref{fig:l131} (right)\footnote{For quiver theories on D3-branes probing non-compact Calabi-Yau spaces, the Newton polygon is the toric diagram of the CY threefold. 
}.

Having discussed how to compute the Newton polygon from the brane tiling, it is natural to ask about the reverse: obtaining the tiling from the Newton polygon. It is indeed possible and with A. Hanany I found a simple procedure \cite{Hanany:2005ss} (see also \cite{Gulotta:2008ef}). It is, however, not a one-to-one mapping because many tilings may correspond to a single Newton polygon. Figure \ref{fig:seiberg} shows operations on the tiling that do not change the Newton polygon\footnote{In the context of 4d $\mathcal{N}=1$ quiver gauge theories, the left operation corresponds to Seiberg duality, while the right operation corresponds to integrating out massive fields.}.

\subsection{Associated integrable systems}

\label{sec:assoc}

An integrable model is associated to a brane tiling as follows \cite{Goncharov2013}. Dynamical variables correspond to oriented edge loops in the tiling. A basis for oriented loops is given by  clockwise loops around each face $f_i$ ($i=1,\ldots,n_F$ where $n_F$ is the number of faces) and two non-trivial loops $f_x$ and $f_y$ winding around the $(1,0)$ and $(0,1)$ cycles of the torus.

The quadratic Poisson brackets between variables in the basis are given by
\bea
  \label{eq:poisson}
  \begin{array}{ccl}
    \{f_i,f_j\} & = & \epsilon_{f_i, f_j} \, f_i f_j \\
    \{f_x,f_y\} & = & (\langle f_x,f_y \rangle + \epsilon_{f_x,f_y}) \, f_x f_y \\
    \{f_x,f_i\} & = & \epsilon_{f_x,f_i} \, f_x f_i \\
    \{f_y,f_i\} & = & \epsilon_{f_y,f_i} \, f_y f_i  \, .
  \end{array}
\eea
Here $\epsilon_{x,y}$ is the number of edges appearing in both loops $x$ and $y$ (counted with orientation) and $\langle f_x,f_y \rangle $ is the intersection number of the two loops.

Note that the face variables are not independent, because they satisfy the relation
\be
  \prod_{i=1}^{n_F} f_i = 1 \, .
\ee

The model can be quantized by considering the q-deformed algebra
\be
  X_i X_j = q^{\{ x_i, x_j \}} X_j X_i \, ,
\ee
where $X_i = e^{x_i}$ are the operators corresponding to the loops and $q=e^{-2\pi i \hbar}$.

Perfect matchings in the tiling can be converted into loops if one superimposes a {\it reference perfect matching}. If an edge appears in both matchings, then it should be removed. The result is a set of non-intersecting loops which is a dynamical variable in the integrable model and thus it can be expressed in terms of the variables in the basis ($f_x, f_y, f_i$). Perfect matchings correspond to lattice points in the Newton polygon. In order to obtain the dynamical variable corresponding to a lattice point, one needs to sum the contribution from all corresponding perfect matchings.

Goncharov and Kenyon showed \cite{Goncharov2013} that the Poisson brackets (or the commutators) define a (0+1)-dimensional (quantum) integrable system. Conserved charges are given by the Hamiltonians and the Casimirs. Hamiltonians commute with each other while Casimirs commute with everything.  Hamiltonians are given by (the contribution of perfect matchings corresponding to) internal lattice points in the Newton polygon. Casimirs are given by ratios of two consecutive lattice points on the boundary of the polygon.

A simple counting of degrees of freedom may convince the reader that the model can indeed satisfy the requirements for Liouville integrability.
Let $I$ denote the number of internal points in the Newton polygon and $B$ denote the number of points on its boundary.
We then have $n_F + 1$ independent dynamical variables in the basis, $I$ Hamiltonians and $B-1$ independent Casimirs. The dimension of the phase space can be reduced by using the Casimirs to express some of the dynamical variables. The reduced dimension is $d= n_F  - B + 2$, which, by Pick's theorem, is equal to $2I$, i.e. twice the number of Hamiltonians. This implies that the system is integrable.

The spectral curve of the integrable model is given by the determinant of the Kasteleyn matrix of the brane tiling,
\be
  P(x,y)\equiv \det K(x,y) = 0 \, .
\ee

In the following we will use this technology to compute the spectral curve of segmented strings.
In order to do this, we need to embed the string dynamics into a cluster integrable model. This will be the topic of the next section.

\clearpage

\begin{figure}[h]
\begin{center}
\includegraphics[width=6cm]{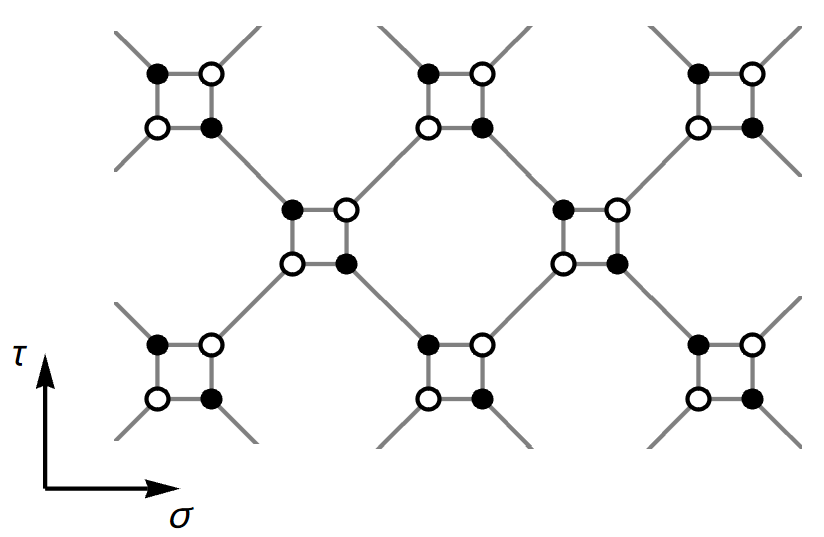}
\caption{\label{fig:onshell} On-shell diagram constructed from a segmented string. Valence four composite vertices in Figure \ref{fig:sub} (right) where left- and right-moving kinks collide have been replaced by four-particle tree-level diagrams (little squares). After collapsing adjacent vertices of the same color, one obtains a square lattice.
}
\end{center}
\end{figure}

\section{Segmented strings and $Y^{n,0}$ brane tilings}

Why are segmented strings related to brane tilings? A hint comes from investigating the structure of the embedding. Kink worldlines are null as depicted in Figure \ref{fig:sub} (right). Thus, one might interpret this graph as an on-shell diagram in the sense of \cite{Arkani-Hamed:2012zlh}.
One might imagine that the collision points are composite vertices which can be replaced by  four-particle tree-level diagrams as in  Figure \ref{fig:onshell}. After collapsing adjacent nodes of the same color, one ends up with a bipartite square lattice. Note, however, that the resulting diagram is not a brane tiling as it is only periodic in the $\sigma$ worldsheet direction. Nevertheless, it suggests that  square lattices may play a role in the description.

\begin{figure}[h]
\begin{center}
\includegraphics[width=3.3cm]{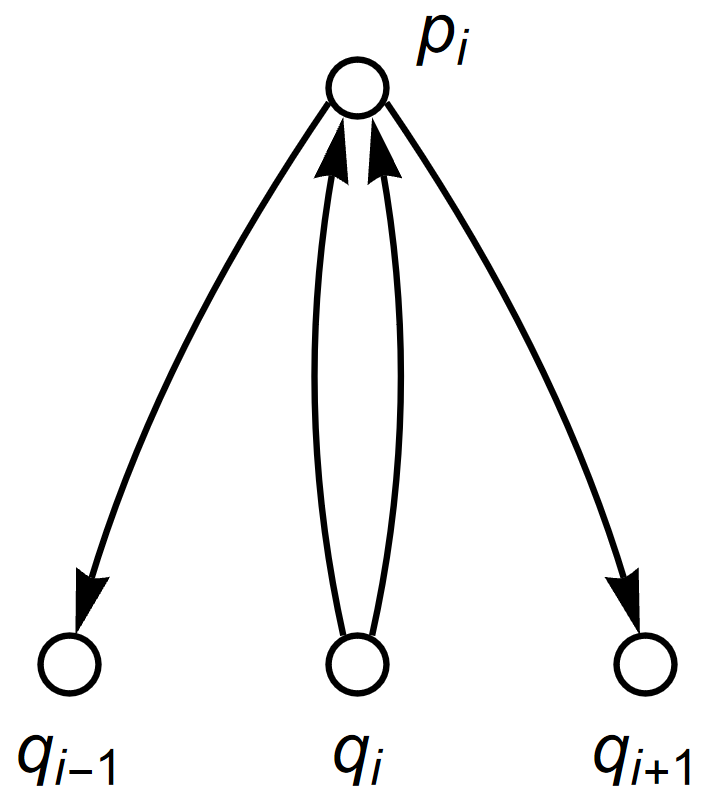}\qquad\qquad\qquad
\includegraphics[width=8cm]{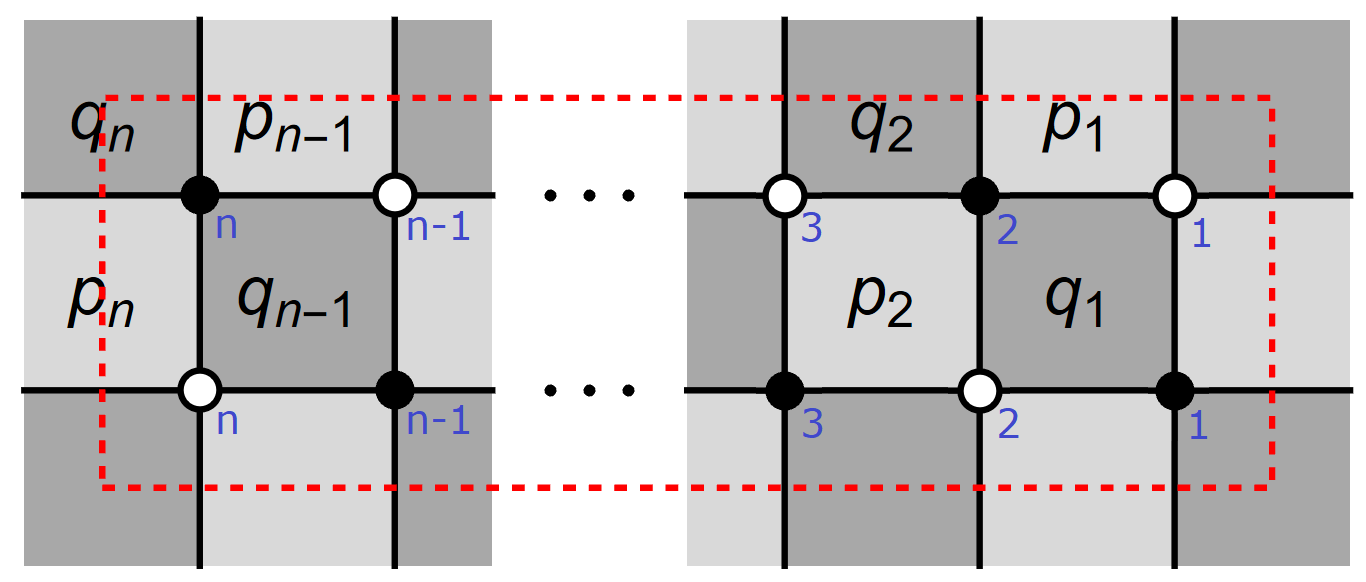}
\caption{\label{fig:yp0}{\it Left:} Local quiver showing the $q$ nodes which are connected to a given $p_i$ node. {\it Right:} Brane tiling corresponding to a segmented string consisting of $n$ segments ($n$ is even). The red dashed line is the boundary of the fundamental cell. The $p$ and $q$ faces have been colored light and dark gray. Blue numbers label the vertices.
}
\end{center}
\end{figure}

After this aside, let us consider the integrable pentagram map \cite{Schwartz}. It associates to a projective polygon a new one given by the intersection points of short diagonals. As shown in \cite{gekhtman2012higher}, a special version, the leapfrog map, satisfies the same discrete equation that governs the time evolution of celestial variables  (\ref{eq:deqn}).
The authors provide the corresponding local quiver which is shown in Figure \ref{fig:yp0} (left). The corresponding brane tiling is easy to find and it is depicted in Figure \ref{fig:yp0} (right). It is in fact the $Y^{n,0}$ tiling\footnote{
Due to its simple structure, one can think of the $Y^{n,0}$ quiver as the `square product' of Dynkin diagrams of the $\hat A_1$ and $\hat A_n$ affine Lie algebras. Generalizations can be considered by replacing these algebras with other ones \cite{Yamazaki:2016aam}.}, which contains $n_V = 2n$ vertices and $n_F = 2n$ square faces labelled alternatingly by $p_i$ and $q_i$ ($i=1,\ldots, n$).
These brane tilings have been analysed in the context of integrable models in \cite{Eager:2011dp}.

\clearpage

\subsection{Loop variables}

We would like to relate the loop variables associated to faces of the tiling to the celestial variables of segmented strings. Let us define the cross-ratio
\be
  (a, \, b; \, c, \, d) \equiv {(a-b)(c-d) \ov (a-d) (b-c)} \, .
\ee
The $p$ and $q$ variables can be computed on the $j^\textrm{th}$ discrete timeslice by \cite{gekhtman2012higher}
\be
  \label{eq:pdef}
  p_{i,j} = (a_{i+1,j}, \, a_{i+1,j+1};  \, a_{i+2,j}, \, a_{i+2, j+1} )
\ee
\be
  \label{eq:qdef}
    q_{i,j} = {(a_{i,j}, \, a_{i+1,j+1};  \, a_{i+2,j+1}, \, a_{i+3, j} )
             (a_{i+1,j}, \, a_{i+2,j};  \, a_{i+2,j+1}, \, a_{i+3, j} ) \ov
             (a_{i,j}, \, a_{i+1,j};  \, a_{i+2,j}, \, a_{i+3, j} )
             (a_{i+1,j}, \, a_{i+1,j+1};  \, a_{i+2,j+1}, \, a_{i+3, j} )
             }
\ee
where $a_{i,j}$ are the celestial fields, see Figure \ref{fig:sub} (right).

\begin{figure}[h]
\begin{center}
\includegraphics[width=6cm]{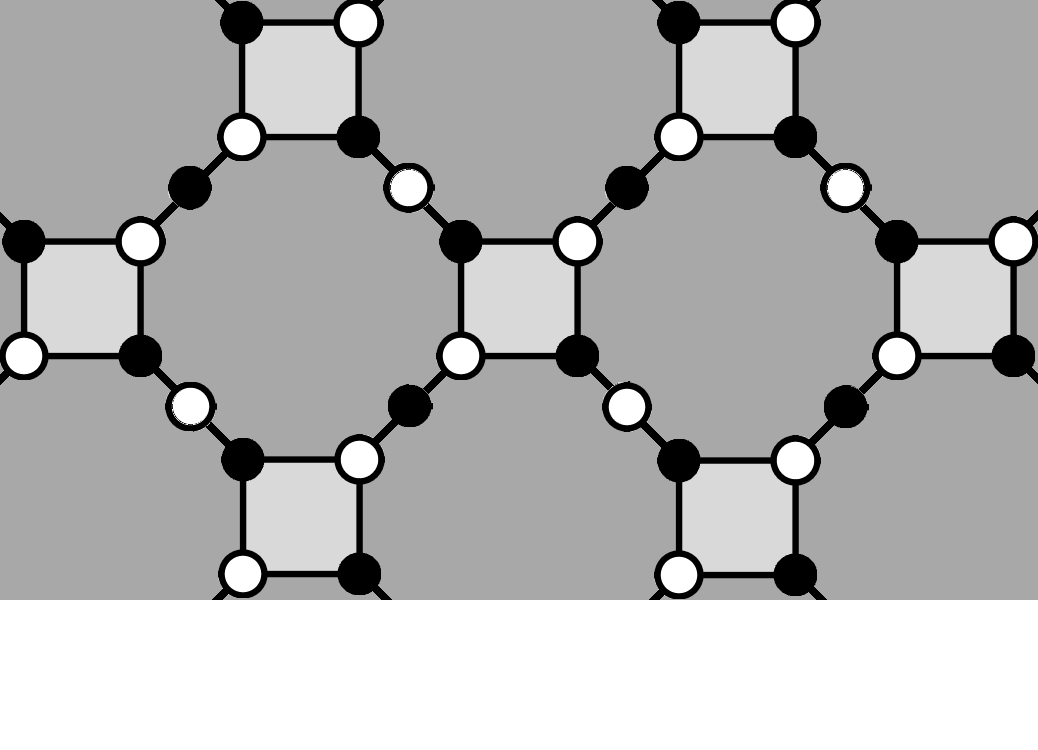}\qquad\qquad\qquad
\includegraphics[width=6cm]{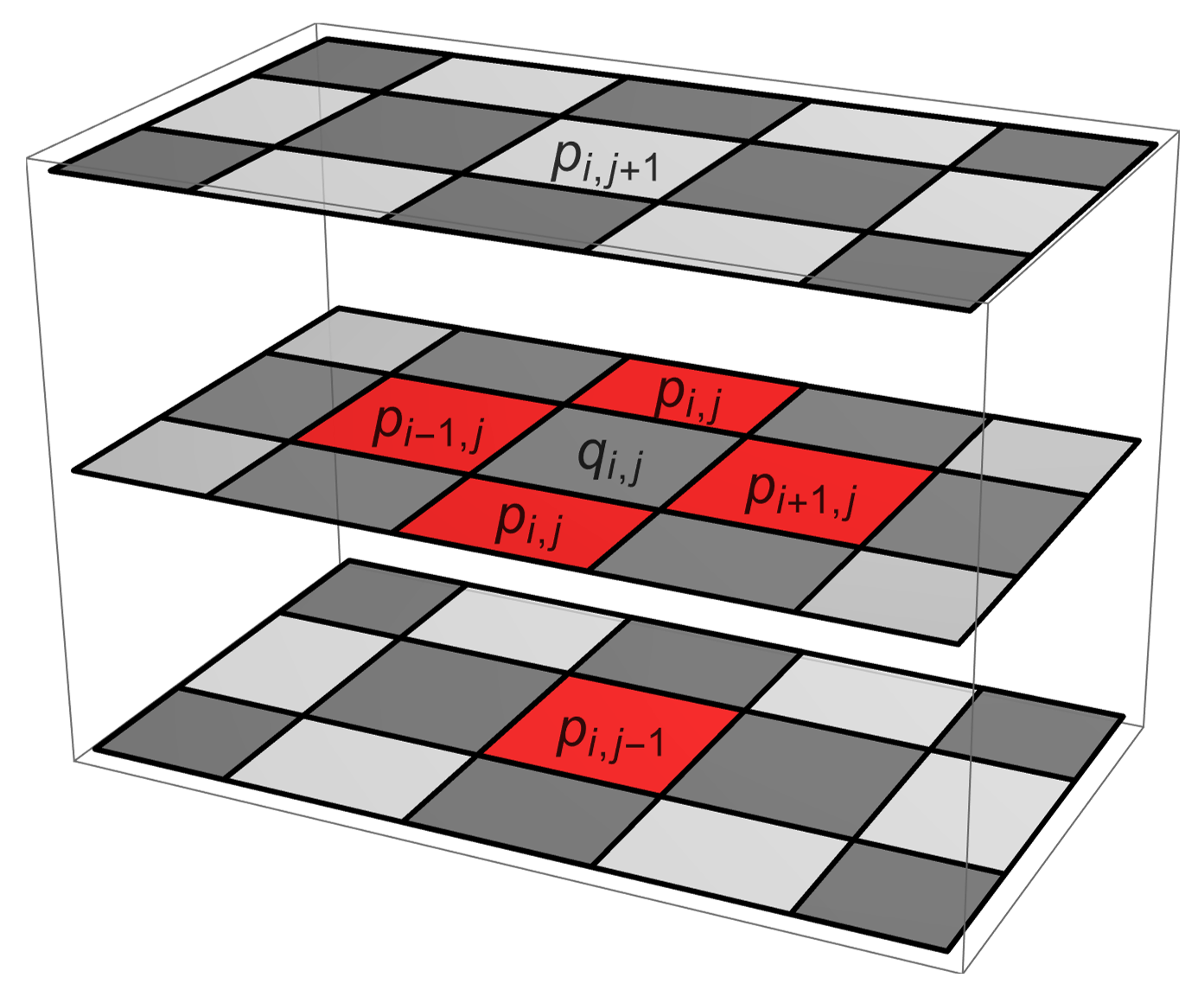}
\caption{\label{fig:seiberg2} {\it Left:} (Partial) brane tiling with cluster transformations performed on the $p$ faces (light gray). Valence two nodes can be removed and the resulting brane tiling is again a square lattice (with flipped vertex colors).  {\it Right:} The value of $p_{i,j+1}$ is determined by the $p$ variables on the red faces on previous layers.
}
\end{center}
\end{figure}

In order to make a discrete step forward in time ($j \rightarrow j+1$), let us consider cluster transformations corresponding to quiver mutations at $p$ nodes.  These transformations commute. After performing all of them simultaneously, the resulting quiver is identical to the original one with $p_i$ and $q_i$ exchanged.  On the brane tiling this is achieved by performing an urban renewal transformation (see Figure \ref{fig:seiberg}  (left)) at every $p$ face. The result of this step is shown in Figure \ref{fig:seiberg2} (left). After urban renewal, valence two vertices can be eliminated by applying the transformation on Figure \ref{fig:seiberg} (right). Note that the color of the vertices has flipped.
Thus, one can compose cluster transformations with exchanging the letters $p \leftrightarrow q$ to arrive at the same quiver. Explicitly, the cluster transformations give
\be
  \label{eq:eomy}
  q_{i,j+1} = {1 \ov p_{i,j}} \, , \qquad
  p _{i,j+1} = q_{i,j}p_{i,j}^2{(1+p_{i-1,j})(1+p_{i+1,j})\ov (1+p_{i,j})^2 } \, .
\ee

\noindent
It is straightforward to check that (\ref{eq:deqn}) implies these equations.

\subsection{The Y-system}

Note that (\ref{eq:eomy}) can be cast in the following form,
\be
  \label{eq:eomy2}
  p _{i,j-1}p _{i,j+1} = {(1+p_{i-1,j})(1+p_{i+1,j})\ov (1+p_{i,j}^{-1})^2 }
\ee
Hence, the face loop variables of the brane tiling form a Y-system\footnote{Not to be confused with the $Y$ of $Y^{p,q}$.} \cite{ZAMOLODCHIKOV1991391} (see \cite{Kuniba:2010ir} for a review). Figure \ref{fig:seiberg2} (right) shows the $j-1, j, j+1$ layers of brane tilings. $p_{i,j+1}$ can be determined from the red faces via the `cube rule' in (\ref{eq:eomy2}). Since $q_{i,j} = p_{i,j-1}^{-1}$, one can think of the brane tiling as a tool for encoding two layers of the Y-system.

Y-systems with fixed (non-periodic) boundary conditions have appeared in AdS/CFT \cite{Gromov:2009tv, Bombardelli:2009ns, Gromov:2009bc, Arutyunov:2009ur, Gromov:2009tq, Gromov:2011cx} where they describe the anomalous dimensions of local single trace operators in planar $\mathcal{N}=4$ supersymmetric Yang-Mills theory.
It has been shown that gluon scattering amplitudes at strong coupling correspond to Euclidean surfaces that end on the AdS boundary \cite{Alday:2007hr}. Y-systems have also been used in this context \cite{Alday:2010vh} and (\ref{eq:eomy2}) can be considered as the Lorentzian analog\footnote{Note that due to the periodicity of the brane tiling, the periodicity conjecture for solutions of Y-systems \cite{ZAMOLODCHIKOV1991391,Kuniba:1992ev,Gliozzi:1994cs}  does not apply.}.

\clearpage

\subsection{The spectral curve}

The spectral curve can be determined by computing the determinant of the dressed $n\times n$ Kasteleyn matrix of the brane tiling. Figure \ref{fig:yp0} (right) shows the labeling of the vertices. In the Kasteleyn matrix, rows are labelled by white vertices, columns are labeled by black vertices. We assign weight labels $w_{ij}$ to the edges as follows: $i$ and $j$ denote the white and black vertices of the edge, respectively, and we put a tilde over $w$ if the edge crosses the upper or lower boundary of the fundamental cell.

Loop variables corresponding to tiling faces are `magnetic fluxes' computed from the edge weights (without signs) as a product with alternating $s=\pm 1$ powers. We give the  loops a clockwise orientation and thus edges also acquire an orentation. The rule is that $s=1$ if the edge starts on a white node, and $s=-1$ if it starts on a black node. For instance, the first face variable on the right of the tiling in Figure \ref{fig:yp0} (right) is $q_1=  w_{11}  w_{21}^{-1} w_{22}   w_{12}^{-1}$.

The Kasteleyn matrix is tri-diagonal (see \cite{Franco:2005rj} or the appendix for a version without weights)
\be
\label{eq:kast}
K = \left( \begin{array}{cccccccc}
x \tilde w_{11}-w_{11} & w_{12} & 0 & 0 & \hdotsfor{3} & y w_{1,n} \\
w_{21} & w_{22}-{\tilde w_{22}\ov x} & w_{23} & 0 & \hdotsfor{3} & 0\\
0 & w_{32} & x \tilde w_{33}- w_{33} &  w_{34} & 0 & \hdotsfor{2} & 0\\
0 & 0 &  w_{43} &  w_{44}-{\tilde w_{44}\ov x} & w_{45} & 0 & \ldots  & 0\\
& \vdots & & & & \ddots & & \\
{w_{n,1} \ov y} & 0 & \hdotsfor{3} & 0 & w_{n,n-1} & w_{n,n}-{\tilde w_{n,n}\ov x}
\end{array} \right)
\ee

In order to determine the weights in the matrix, let us first define the M\" obius-invariant quantities
\be
  \label{eq:xdef}
  x_{i,j} \equiv -(a_{i+1,j+1}, \, a_{i+2,j};  \, a_{i+1,j}, \, a_{i, j} ) \, ,
\ee
\be
  \label{eq:ydef}
  y_{i,j}\equiv {(a_{i+1,j}-a_{i+1,j+1})(a_{i+2,j}-a_{i+2,j+1})(a_{i,j}-a_{i+1,j}) \ov
  (a_{i+1,j}-a_{i+2,j+1})(a_{i,j}-a_{i+1,j+1})(a_{i+1,j}-a_{i+2,j}) } \, .
\ee
These new quantities are related to the face loop variables via
\be
   \nonumber
   p_i = {y_i \ov x_i} \, , \qquad q_i = {x_{i+1} \ov y_i} \, ,
\ee
where we have suppressed the $j$ indices.

\clearpage

\begin{figure}[h]
\begin{center}
\includegraphics[width=5cm]{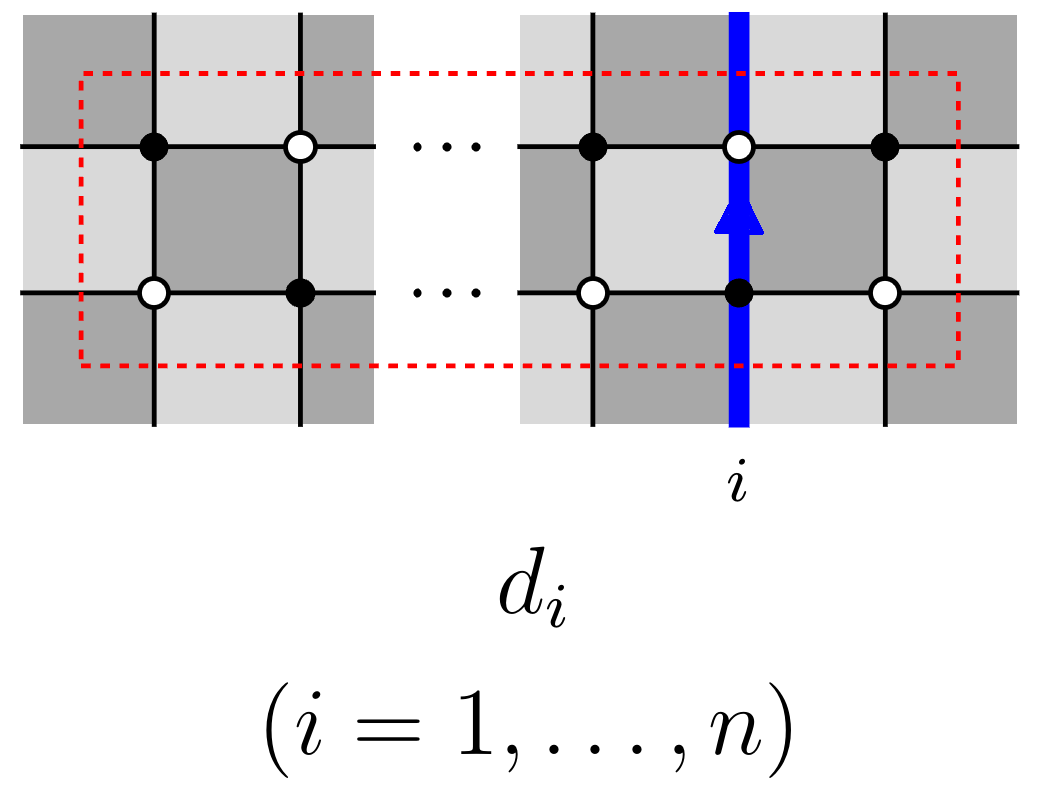} \quad
\includegraphics[width=5cm]{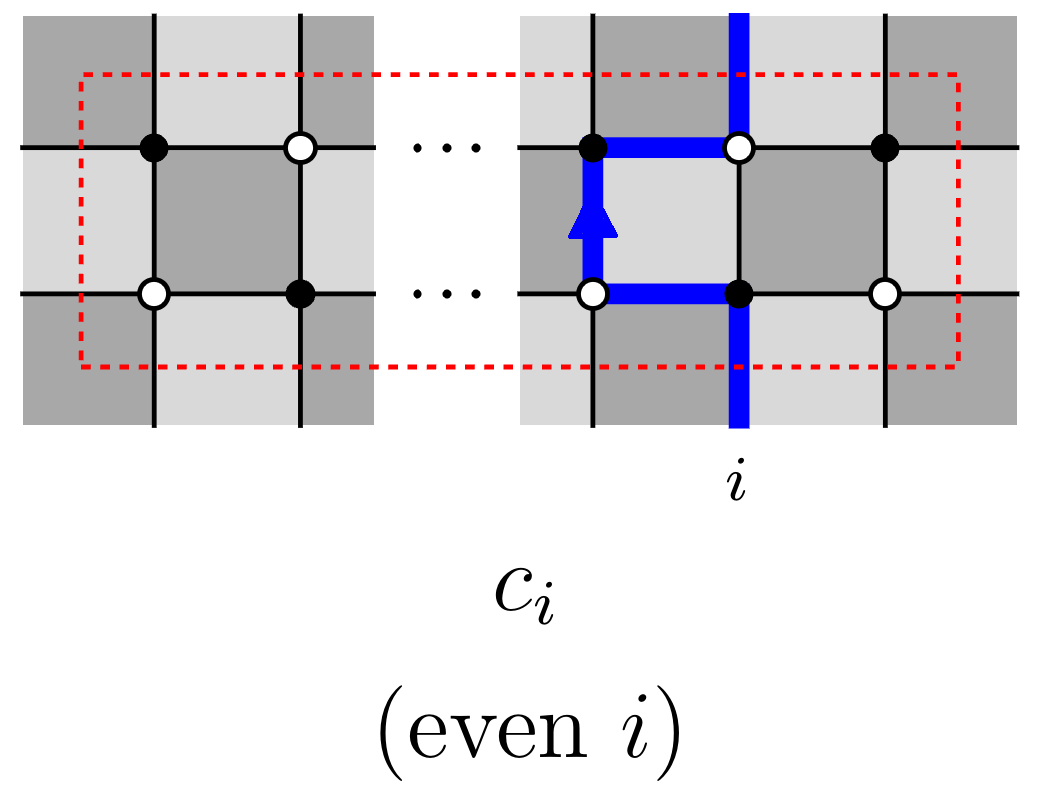} \quad
\includegraphics[width=5cm]{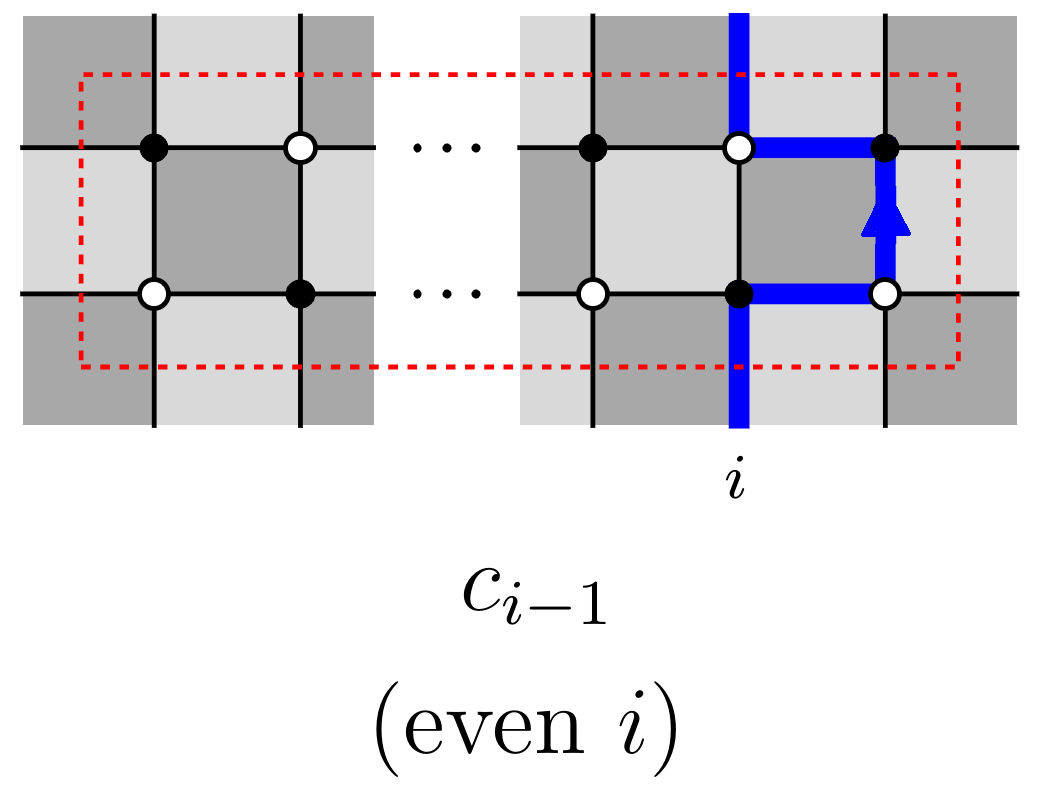}
\caption{\label{fig:di}
A convenient basis for the loop variables is given by the $2n$ cycles $c_i, d_j$ where $i$ labels the vertical lines in the fundamental cell from right to left ($i=1,\ldots, n$).
}
\end{center}
\end{figure}

In fact, $x_i$ and $y_i$ are part of the set of dynamical variables in the integrable model and can   simply be expressed in terms of  the $2n$ loops defined in Figure \ref{fig:di} by noting that
\be
  \nonumber
  x_i = - d_i \, , \qquad y_i = -c_i \, .
\ee

We assign the following weights to the edges so that the magnetic fluxes give the correct face variables,
\be
  \nonumber
  w_{i,i}  =\left\{\begin{array}{ll}
 x_{i}^{-\half} \quad &  \textrm{ if $i$ is odd}  \\
 x_{i}^{\half} \quad &  \textrm{ if $i$ is even}
 \end{array}\right. \, ,
\ee
\be
  \label{eq:edgeweights}
 \tilde w_{i,i}  =\left\{\begin{array}{ll}
 -x_{i}^{\half} \quad &  \textrm{ if $i$ is odd}  \\
-x_{i}^{-\half} \quad &  \textrm{ if $i$ is even}
 \end{array}\right. \, ,
\ee
\be
   \nonumber
 w_{i,i+1}  = w_{i+1,i}  = {y_i^{\half} \ov (x_i x_{i+1})^{1\ov 4}}  \, ,
\ee
where $x_{n+1} \equiv x_1$.

Note that other edge weights are possible. Different weight assignments are connected by gauge transformations.  A local gauge transformation at a node multiplies the weights of all edges incident to that node by a common constant. Weight systems related by gauge transformations are equivalent since loop variables are gauge invariants.
There are $4n$ edges and $2n-1$ independent gauge transformations. Thus, without further restrictions, the number of independent dynamical variables is $n_F+1$.

\clearpage

\begin{figure}[h]
\begin{center}
\includegraphics[width=6cm]{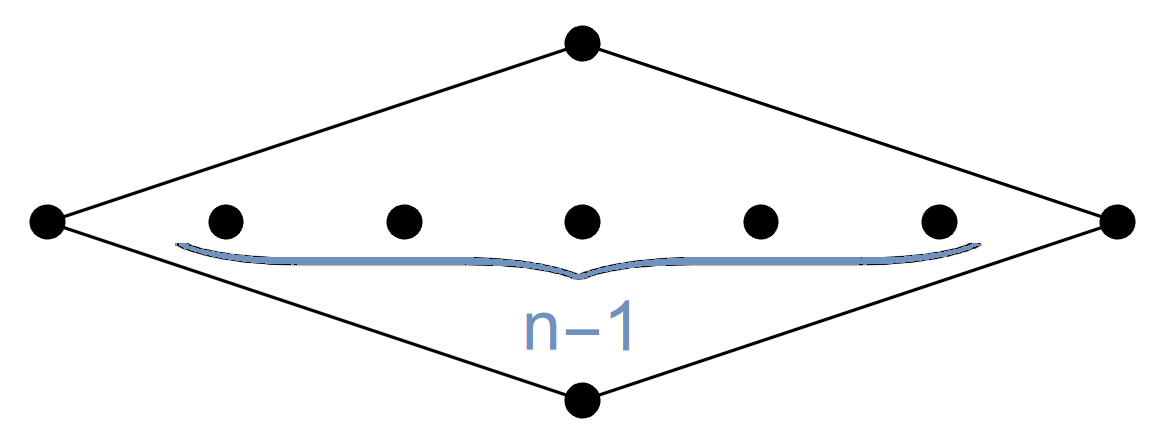}
\caption{\label{fig:toric} The Newton polygon of the spectral curve for a string with $n$ segments. The Hamiltonians corresponding to the $n-1$ internal points are not all independent due to constraints which ensure that the string closes in target space. The system has one invariant Casimir.
}
\end{center}
\end{figure}

Finally, the spectral curve is given by
\be
  \nonumber
  \det K = 0 \, .
\ee

The Newton polygon\footnote{Incidentally, this is the toric diagram of the $Y^{n,0}$ geometry, which is an orbifold of the conifold.} for a generic $n$ value is shown in Figure \ref{fig:toric}. There are $B-1 = 3$ Casimirs. It is important to note that not all Hamiltonians are independent. This is due to the fact that the string has to form a closed loop in AdS$_3$, which imposes certain constraints on the celestial variables and on the variables of the integrable model.

\clearpage

\begin{figure}[h]
\begin{center}
\includegraphics[width=3.5cm]{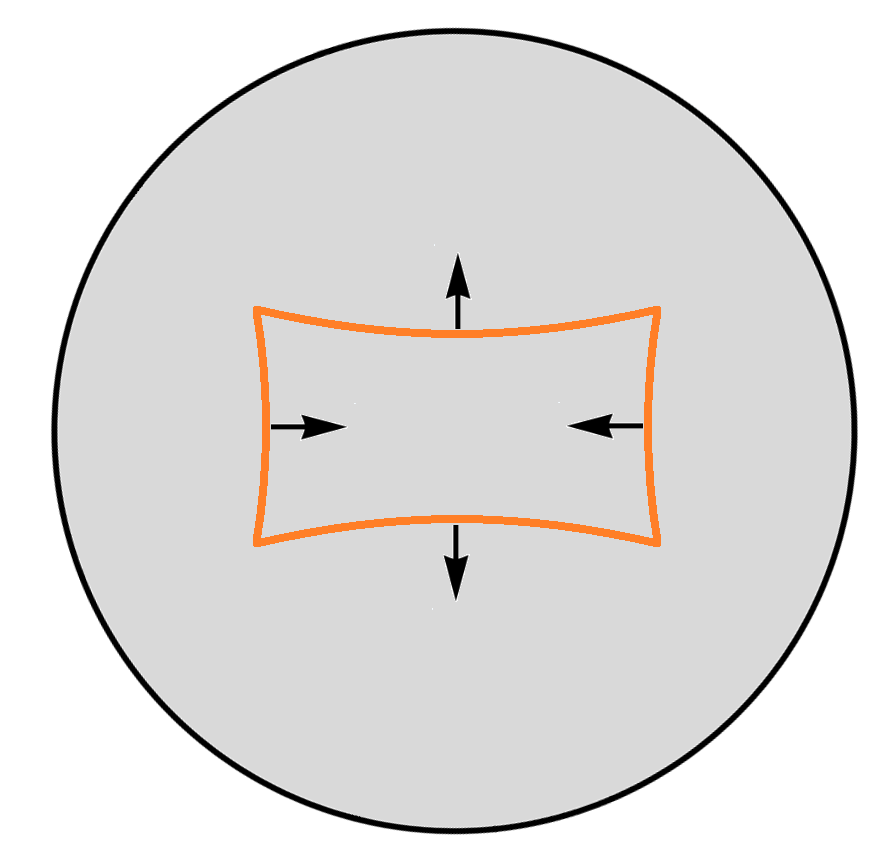} \qquad\qquad
\includegraphics[width=6.6cm]{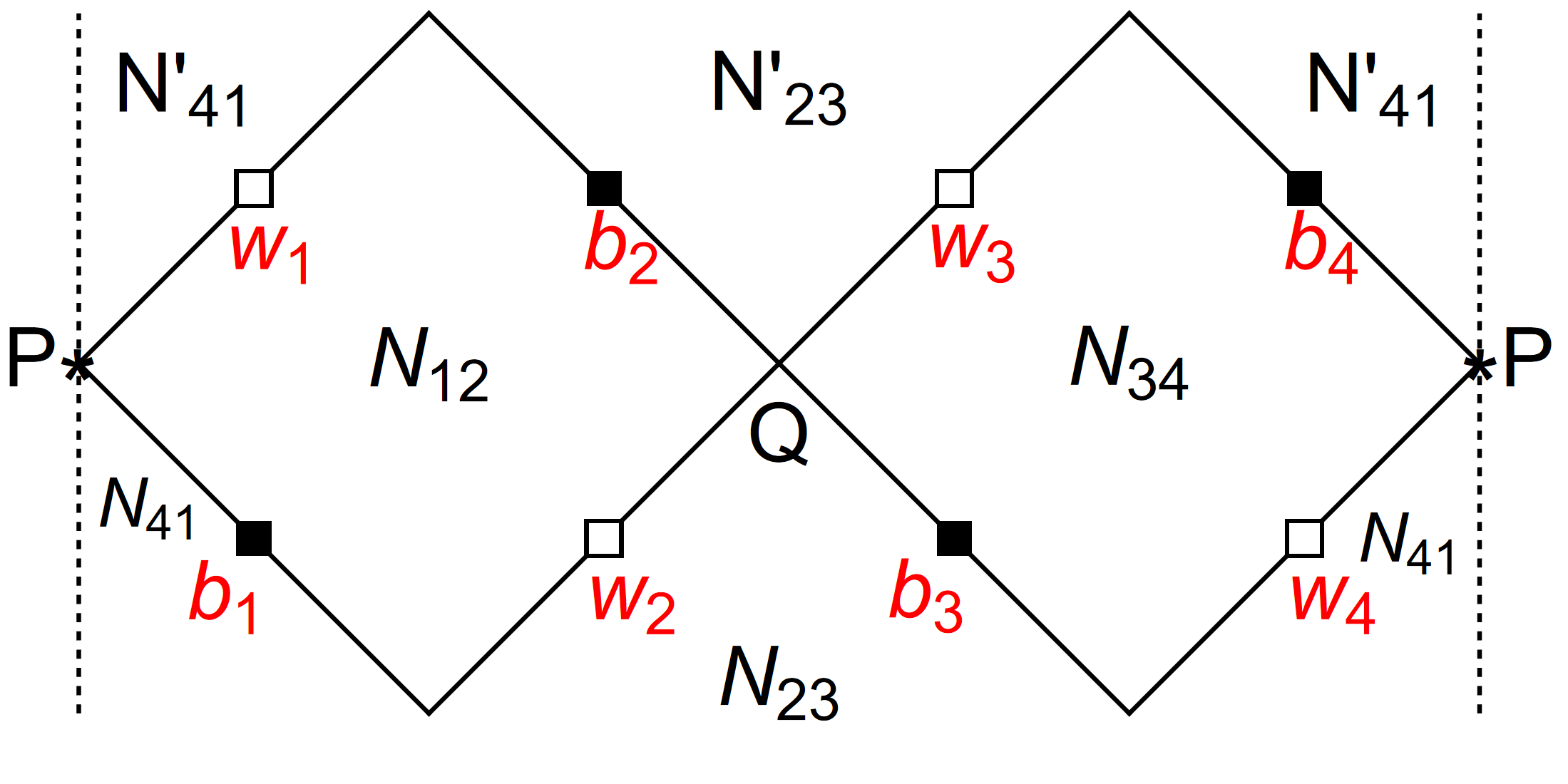}
\caption{\label{fig:4seg} {\it Left:} Segmented string with four kinks on a timeslice of global AdS$_3$. In the applied coordinate system the segments are circular arcs. Arrows indicate the direction of motion of the segments. See \cite{anim} for an animation made by the authors of \cite{Callebaut:2015fsa}. {\it Right:} String worldsheet to be glued along the dashed lines. $N_{12}, N_{23}, N'_{23}, N_{34}, N_{41}, N'_{41}$ label the normal vectors of the corresponding AdS$_2$ patches. $b_i$ and $w_j$ are celestial variables. $P$ and $Q$ are spacetime points.
}
\end{center}
\end{figure}

\begin{figure}[h]
\begin{center}
\includegraphics[width=6cm]{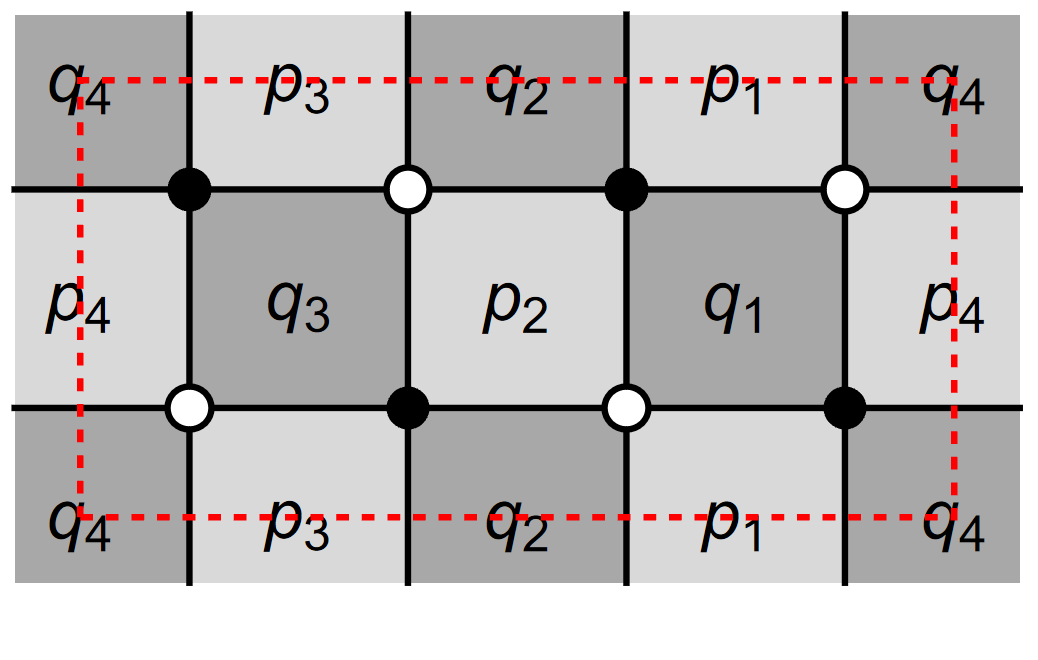}\qquad\qquad\qquad
\includegraphics[width=7cm]{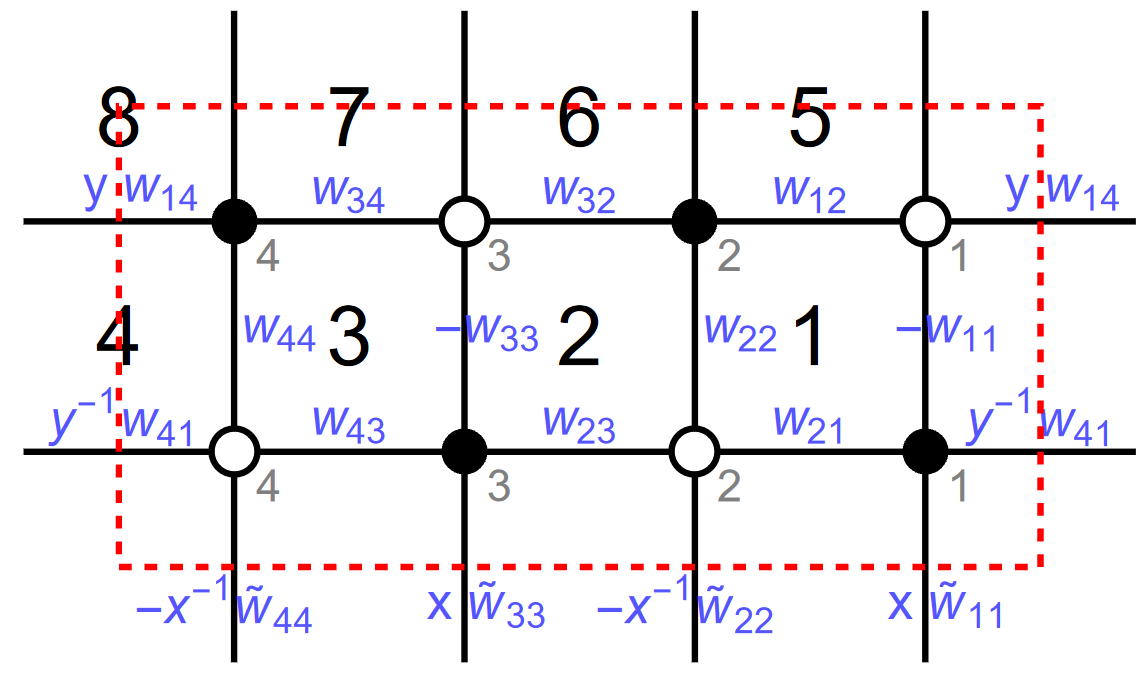}
\caption{\label{fig:y40}  {\it Left:} Brane tiling for a string with four segments.
 {\it Right:} Weights associated to edges in the tiling.
}
\end{center}
\end{figure}

\section{An example with four segments}
\label{sec:ex4}

This section explains how to assign weights to edges in the brane tiling through the example of a string consisting of four segments (see Figure \ref{fig:4seg} (left) for the embedding). This will allow us to express the spectral curve directly in terms of celestial variables, depicted in Figure \ref{fig:4seg} (right).

The brane tiling for the string is shown in Figure \ref{fig:y40} (left). The $p$ and $q$ variables are indicated on the faces. To make contact with section \ref{sec:assoc}, we will also use the $f_i$ notation for face loop variables ($i=1,\ldots, 8$). These labels along with the edge weights are shown in Figure \ref{fig:y40} (right).

\begin{figure}[h]
\begin{center}
\includegraphics[width=5cm]{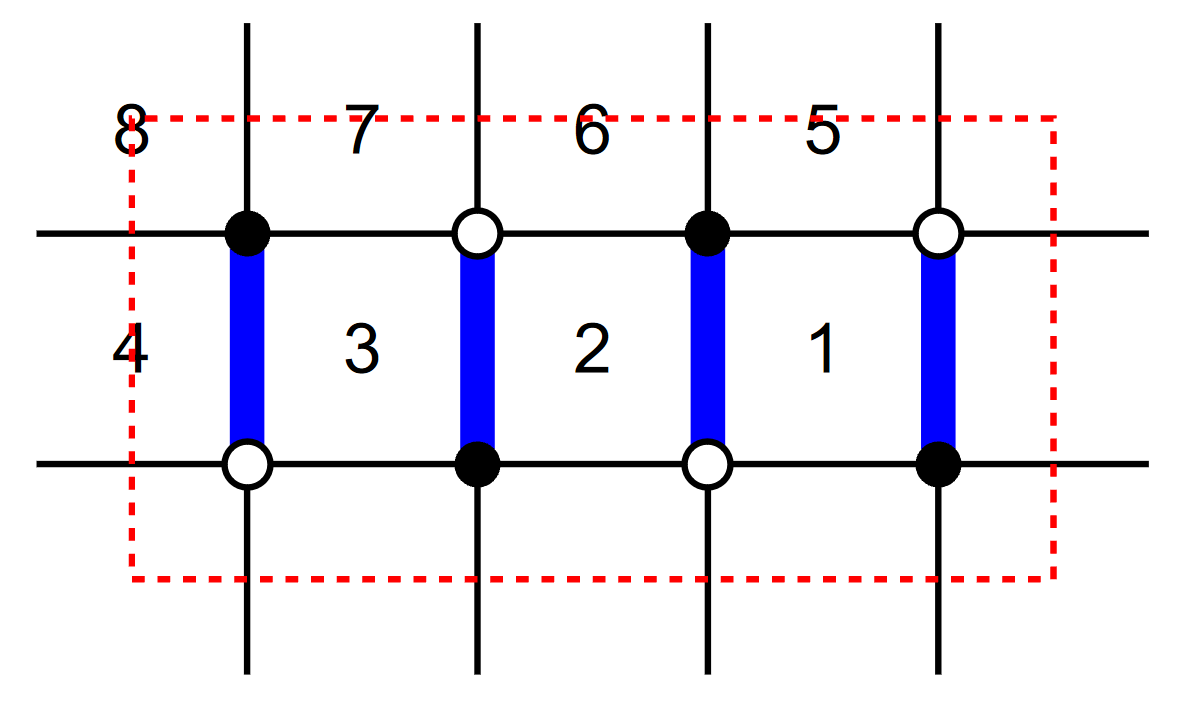} \qquad
\includegraphics[width=5cm]{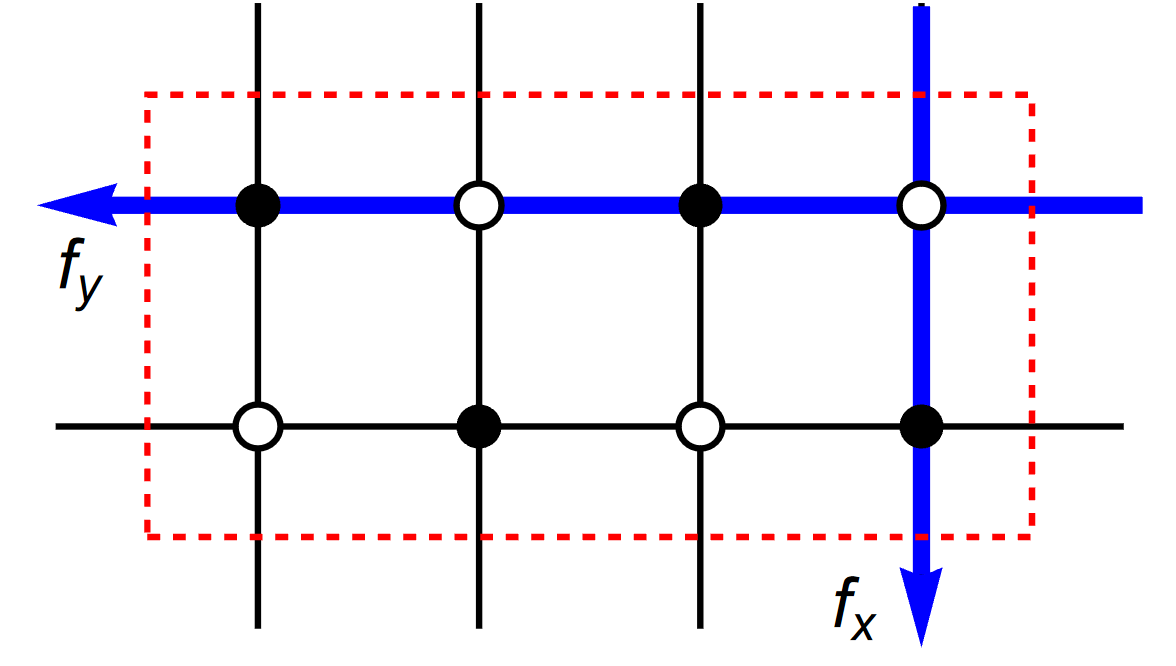}
\caption{\label{fig:y40ref} {\it Left:} Reference perfect matching. Perfect matchings are listed in the appendix.
{\it Right:} Two non-trivial cycles in the tiling.
}
\end{center}
\end{figure}

We have
\be
  \nonumber
     f_1 \equiv q_1 \quad
     f_2 \equiv p_2  \quad
     f_3 \equiv q_3  \quad
     f_4 \equiv p_4  \quad
     f_5 \equiv p_1 \quad
     f_6 \equiv q_2  \quad
     f_7 \equiv p_3  \quad
     f_8 \equiv q_4 \, .
\ee
The loop variables are `magnetic fluxes' computed from the edge weights,
\bea
  \nonumber
 && f_1 = {w_{11} w_{22} \ov w_{21} w_{12}} \qquad
  f_2 = {w_{23}  w_{32} \ov w_{33} w_{22}}  \qquad
  f_3 = {w_{33} w_{44} \ov w_{43}  w_{34}} \qquad
  f_4 = {w_{41} w_{14} \ov w_{11} w_{44}} \\
  \label{eq:faceweight}
 &&
 %\qquad\qquad
 f_5 = {w_{21} w_{12} \ov \tilde w_{22} \tilde w_{11}}  \qquad
  f_6 = {\tilde w_{22} \tilde w_{33} \ov w_{32} w_{23}}  \qquad
  f_7 = {w_{43}  w_{34} \ov \tilde w_{33} \tilde w_{44}} \qquad
  f_8 = {\tilde w_{11} \tilde w_{44} \ov w_{14} w_{41}} \, ,
\eea
and we also have a relation between them,
\be
  \nonumber
  \prod_{i=1}^{8} f_i = 1 \, .
\ee
The basis of dynamical variables in the integrable model also contain two non-trivial loops $f_x, f_y$ as shown in Figure \ref{fig:y40ref} (right), which are given in terms of the weights as
\be
  \nonumber
  f_x = {w_{11} \ov \tilde w_{11}}  \qquad
   f_y = {w_{12} w_{34} \ov w_{32} w_{14}} \, .
\ee

\subsection{The spectral curve}

The Kasteleyn matrix,
\be
K = \left( \begin{array}{cccc}
x \tilde w_{11}-w_{11} & w_{12} & 0 &  y w_{14} \\
w_{21} & w_{22}-x^{-1}\tilde w_{22} & w_{23} & 0 \\
0 & w_{32} & x \tilde w_{33}-w_{33} & w_{34} \\
y^{-1}w_{41} & 0 & w_{43} & w_{44}-x^{-1}\tilde w_{44}
\end{array} \right)
\ee

We divide the determinant of the Kasteleyn matrix by the contribution of a reference matching in Figure \ref{fig:y40ref} (left):
\be
  P( x, y) = {\det K  \ov w_{11} w_{22} w_{33} w_{44}}
\ee

\begin{figure}[h]
\begin{center}
\includegraphics[width=4cm]{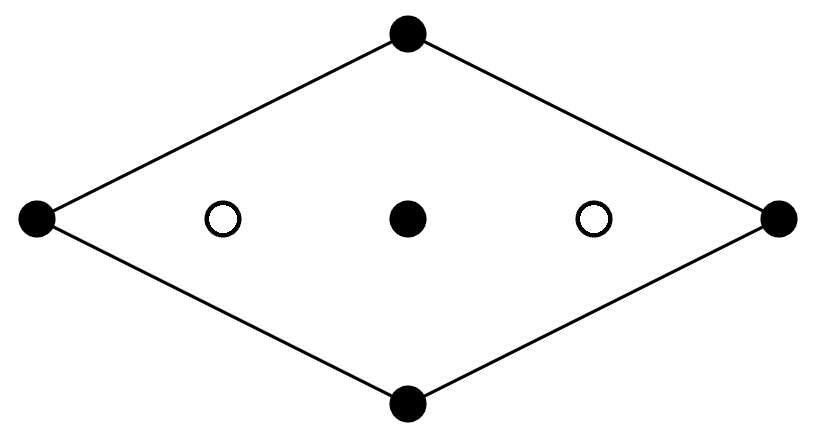}
\caption{\label{fig:toric4} Newton polygon of the spectral curve for a string with four segments. The middle point is at $(0,0)$. The other two internal lattice points at $(\pm 1,0)$ are shown as empty circles, indicating that the coefficients of the corresponding terms in the spectral curve vanish. This is a consequence of the constraints which ensure that the string closes in AdS$_3$.
}
\end{center}
\end{figure}

The Kasteleyn matrix depends on the weights $x$ and $y$ which we rescale by introducing
\be
  \tilde x =  x / f_x \qquad \textrm{and} \qquad \tilde y =   y / f_y
\ee
The resulting polynomial can be written in terms of the $f_i$ loop variables,
\bea
\nonumber
P(\tilde x,\tilde y)   &&
\hskip -0.5cm = 1+f_2+f_4+{f_1}^{-1}+{f_3}^{-1}+f_2 f_4+f_2 f_6+f_4 f_8+(f_3 f_7)^{-1}+({f_1 f_3})^{-1}
\\
\nonumber
&&
\hskip -0.5cm +
({f_1 f_5})^{-1}  +f_2 f_4 f_6+f_2 f_4 f_8  +
({f_1 f_3 f_5})^{-1}+({f_1 f_3 f_7})^{-1}  +f_2 f_4 f_6 f_8
\\
\nonumber
&&
\hskip -0.5cm -
\tilde x \Big[1+f_2+{f_3}^{-1}+f_2 f_6+({f_3 f_7})^{-1}+f_2 f_5 f_6 +f_1 f_2 f_5 f_6+f_1 f_2 f_4 f_5 f_6\Big]
\\
\nonumber
&& \hskip -0.5cm  -
\tilde x^{-1}\Big[f_4 f_8 +({f_1 f_5})^{-1} +f_2 f_4 f_8+{f_4 f_8}{f_1}^{-1} +({f_1 f_3 f_5})^{-1}+{f_4}({f_1 f_5})^{-1} +f_2 f_4 f_6 f_8
\\
\label{eq:spec4}
&& \hskip -0.5cm +{f_4 f_8}({f_1 f_5})^{-1} \Big] +
\tilde x^2 f_1 f_2 f_5 f_6
+
\tilde x^{-2}{f_4 f_8}({f_1 f_5})^{-1}
-
\tilde y ({f_1 f_3})^{-1}
-
\tilde y^{-1}{f_2 f_4}
\eea

Its Newton polygon is shown in Figure \ref{fig:toric4}.
On the initial timeslice ($j=1$) we have $a_{1,1} = b_1$, $a_{1,2} = w_1$, $a_{2,1} = w_2$, $a_{2,2} = b_2$, etc. Plugging these into (\ref{eq:qdef}) and  (\ref{eq:pdef}) one can compute the loop variables from the celestial variables,
\bea
  \nonumber
   q_1  \equiv f_1  &=& {(b_1 - b_2)(w_2-b_3)^2 (w_3 - w_4) \ov (b_1-w_2)(w_2-b_2)(b_3-w_3)(b_3-w_4)} \\
  \nonumber
  q_2 \equiv f_6 &=& {(b_1 - b_4)(w_2-w_3) (b_3 - w_4)^2 \ov (w_2-b_3)(b_3-w_3)(b_1-w_4)(w_4-b_4)} \\
  \nonumber
  q_3 \equiv f_3 &=& {(b_3 - b_4)(w_1-w_2) (b_1 - w_4)^2 \ov (b_1-w_1)(b_1-w_2)(b_3-w_4)(w_4-b_4)} \\
  \nonumber
  q_4  \equiv f_8 &=& {(b_2 - b_3)(b_1-w_2)^2 (w_1 - w_4) \ov (b_1-w_1)(w_2-b_2)(w_2-b_3)(b_1-w_4)}
\eea
\be
  p_1 \equiv f_5 = {(w_2 - b_2)  (b_3 - w_3) \ov (b_2-b_3)(w_2-w_3) } \, , \qquad
  p_2 \equiv f_2 = {(b_3 - w_3)  (w_4 - b_4) \ov (b_3-b_4)(w_3-w_4) }
\ee
\be
  p_3 \equiv f_7 = {(b_1 - w_1)  (w_4 - b_4) \ov (b_1-b_4)(w_1-w_4) }\, , \qquad
  p_4 \equiv f_4 = {(b_1 - w_1)  (w_2 - b_2) \ov (b_1-b_2)(w_1-w_2) }
\ee
Plugging these expressions into (\ref{eq:spec4}) one obtains the spectral curve in terms of celestial variables,
\be
  \label{eq:spec4a}
  P(\tilde x,\tilde y)  =0 \, .
\ee
This result can be compared to the direct calculation of section \ref{sec:spectral}. The $2\times 2$ Lax monodromy matrix is given by the product,
\be
  \label{eq:oms4}
  \Omega(\zeta) = \Omega_{b_4, w_4}^{-1} \Omega_{b_{3}, w_{3}} \Omega_{b_2, w_2}^{-1} \Omega_{b_1, w_1} \, .
\ee
The spectral curve is then given by
\be
  \label{eq:spec4b}
  Q(\lambda, \zeta) =   \det\le( \lambda \mathbb{1} - \Omega(\zeta) \ri) = 0 \, .
\ee
It is straightforward to check that by identifying
\be
   \tilde x \sim \zeta^2 \, , \qquad \tilde y \sim \lambda \, ,
\ee
the two curves (\ref{eq:spec4a}) and (\ref{eq:spec4b}) are indeed equivalent.

\subsection{Closing constraints}

Finally, note that the celestial variables are not completely independent. The string has to form a closed loop in target space. This condition is equivalent to $\Omega$ in (\ref{eq:oms4}) satisfying $\Omega(i) =   \mathbb{1}$ (as in  (\ref{eq:closeo})), which imposes three constraints on the celestial variables.
Using the constraints, one can solve for three of the celestial variables, e.g. we can express $w_3, b_4, w_4$,
\bea
  \nonumber
  w_3 &=& {b_1 w_1(b_3-w_2)+b_2 w_2(w_1-b_3)+b_1 b_2(w_2-w_1) \ov b_1(b_3-w_1)+b_3(w_1-w_2)+b_2(w_2-b_3) }  \, , \\
  \nonumber
  b_4 &=& {b_1 b_2(b_3+w_1-w_2) + w_1 w_2(b_3-b_2) + b_1 w_1(w_2-2 b_3) \ov
  b_1(b_2-b_3)- w_1 b_3 + b_2(b_3-2 w_2)+w_2(w_1+b_3)} \, , \\
  \nonumber
  w_4 &=& {w_1 b_2(b_3-w_2)+ b_1(b_2(w_1-w_2)+w_2(w_1+b_3)-2 w_1 b_3) \ov b_2(w_1+b_3-2 w_2) + b_1(w_2-b_3)+b_3(w_2-w_1)} \, .
\eea
If these expressions are plugged back into the spectral curve, then the coefficients of $\tilde x $ and $\tilde x^{-1}$  in (\ref{eq:spec4}) vanish. This is indicated in Figure \ref{fig:toric4} by two empty circles inside the Newton polygon.

The integrable model defined in section \ref{sec:assoc} has a six-dimensional phase space, corresponding to twice the number of internal points in the Newton polygon. We have seen that the celestial variables live in a 5d manifold. AdS$_3$ isometries do not change the $p$ and $q$ loop variables in the tiling, because they have been defined via cross-ratios. Isometries act on the celestial variables by M\" obius transformations which can be parametrized by three real numbers. Thus, the integrable model has a two-dimensional reduced phase space (corresponding to the single non-empty circle in the middle of the Newton polygon in Figure \ref{fig:toric4}).

\section{Invariant $2\times 2$ Lax matrices}
\label{sec:con}

In the previous sections we have seen how to embed the dynamics of segmented strings into the cluster transformation dynamics of $Y^{n,0}$ brane tilings. The embedding is not surjective because both left- and right-handed celestial fields on a closed string must be periodic, which imposes certain constraints on the tiling loop variables. The integrable model {\it without} these constrains is the relativistic version \cite{ruijse1990} of the celebrated Toda lattice \cite{toda1, toda2}. It corresponds to a string that has an $SO(2,2)$ monodromy in target space as $\sigma \rightarrow \sigma + 2\pi$. At the level of celestial fields this means that celestial variables are not periodic: they suffer a monodromy (a M\"obius transformation).

In this section we derive (some of) the constraints that the tiling loop variables have to satisfy in order for the corresponding string embedding to form a closed loop. For a string with $n$ segments, there are $2n$ celestial variables.  The closing constraints (\ref{eq:closeo}) give three relations between them. Finally, loop variables are invariant under target space isometries, which act on the celestial fields via M\" obius transformations. Therefore the dimension of the constrained phase space should be $2n-6$, which is less than that of the unconstrained one, i.e. $2I = 2n-2$.

Let us consider a string consisting of $n$ segments (see Figure \ref{fig:spectral}). Its target space embedding closes if the monodromy
\be
%  \label{eq:monodromy}
\Omega(\zeta) = \Omega_{b_n, w_n}^{-1} \Omega_{b_{n-1}, w_{n-1}} \cdots \Omega_{b_2, w_2}^{-1} \Omega_{b_1, w_1} \,
\ee
at $\zeta = i$ is trivial. This is the closing constraint in terms of covariant celestial variables.
In the following, we will express the constraints in terms of $SL(2)$-invariant loop variables.

\begin{figure}[h]
\begin{center}
\includegraphics[width=9cm]{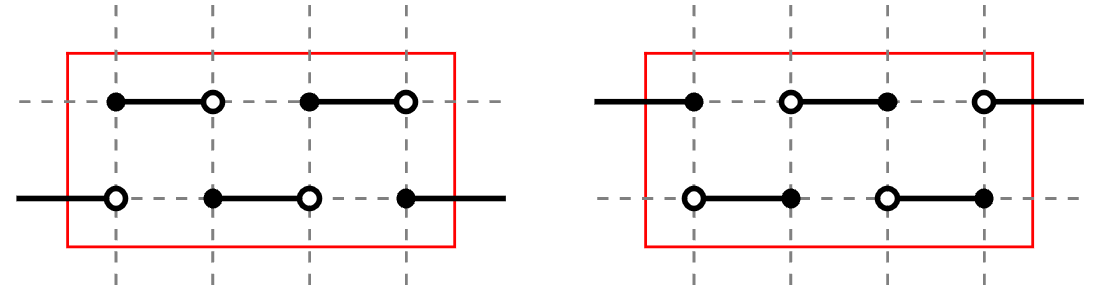}
\caption{\label{fig:fig_refms} Reference perfect matchings corresponding to the $(0,-1)$  and $(0,1)$ (bottom and top) external points in Figure \ref{fig:toric4}.
}
\end{center}
\end{figure}

Let us consider the $n\times n$ Kasteleyn matrix. It is an $SL(2)$ invariant, because its entries, the edge weights (\ref{eq:edgeweights}), contain invariant combinations of the celestial fields, namely the $x_i, y_i$ quantities which correspond to the $c_i, d_i$ loops in Figure \ref{fig:di}. The determinant of the Kasteleyn matrix can be rewritten using the following identity \cite{Eager:2011dp}
\be
\det
\begin{pmatrix}
A_1 & B_1 & & C_0 \\
C_1 & \ddots & \ddots & \\
 & \ddots & \ddots & B_{n-1} \\
 B_0 & & C_{n-1} & A_n
\end{pmatrix}
=
(-1)^{n-1}\left( \prod_{j=0}^n B_j + \prod_{j=0}^n C_j \right) + \Tr L_n L_{n-1} \dots L_1
\ee
where
\be
L_j \equiv \begin{pmatrix}
A_j & - B_{j-1} C_{j-1} \\
1 & 0
\end{pmatrix} \, ,
\ee
is an invariant $2\times 2$ Lax matrix.
If we plug in  the matrix entries of (\ref{eq:kast}), then we get
\bea
  \nonumber
  L_1=\left( \begin{array}{cc}
  x \tilde w_{11}-w_{11} & \  -w_{1,n}w_{n,1}  \\
  1 & 0
  \end{array}\right) \, , & &
  L_2=\left( \begin{array}{cc}
  w_{22}-{ \tilde w_{22} \ov x} & \  -w_{12}w_{21}  \\
  1 & 0
  \end{array}\right) \, , \\
  \nonumber
   L_3=\left( \begin{array}{cc}
  x \tilde w_{33}-w_{33} & \  -w_{23}w_{32}  \\
  1 & 0
  \end{array}\right) \, , & &
  L_4=\left( \begin{array}{cc}
  w_{44}-{ \tilde w_{44} \ov x}  & \  -w_{34}w_{43}  \\
  1 & 0
  \end{array}\right) \, , \ldots \ \textrm{etc.}
\eea
These matrices only contain the $x_i, y_i$ dynamical loop variables of the integrable model.

\subsection{Constrained loop variables}

We have
\be
  \det K =
 (y+y^{-1}) \underbrace{\prod_{j=0}^n  \le({y_i \ov x_i} \ri)^{\half}}_{K}  + \Tr L_n L_{n-1} \dots L_1 \, .
\ee
$K$ is nothing but the contribution of either of the two reference matchings corresponding to the $(0,-1)$  and $(0,1)$ (bottom and top) external points in the Newton polygon,
\be
  \nonumber
  K = w_{12} w_{23} \cdots w_{n-1,n} w_{n,1} = w_{21} w_{32} \cdots  w_{n,n-1} w_{1,n} \, .
\ee
For $n=4$ these perfect matchings are shown in Figure~\ref{fig:fig_refms}.

\begin{figure}[h]
\begin{center}
\includegraphics[width=9.5cm]{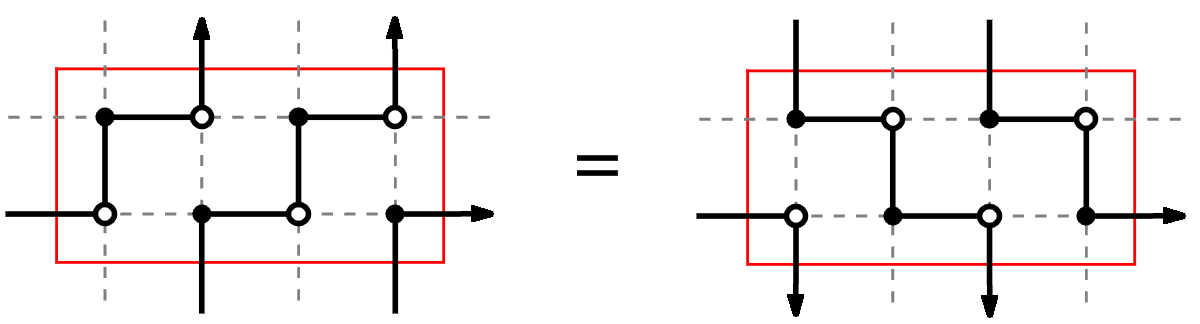}
\caption{\label{fig:closingeq1} A particularly simple closing constraint that equates two loop variables.
}
\end{center}
\end{figure}

Instead of $\Omega$ we can now consider the  matrix
\be
   T \equiv  L_N L_{N-1} \dots L_1 \, .
\ee
Its trace computes the quasimomentum $p$
\be
  \cos p = -{1 \ov 2K } \Tr T  = \half \Tr \Omega \, .
\ee
At the special values of the spectral parameter $x =  \pm 1$ ($\zeta = 1,i$),  the quasimomentum should behave up to linear order as
\bea
  \label{eq:lin1}
 &&  \cos p = 1 +  \mathcal{O}\le[(x- 1)^2 \ri] \quad \textrm{as} \quad x \to 1 \\
  \label{eq:lin2}
  &&  \cos p = 1 +  \mathcal{O}\le[(x+ 1)^2 \ri] \quad \textrm{as} \quad x \to -1 \, .
\eea
These equations give the desired constraints which can be expressed in terms of the loop variables of the integrable model.

\begin{figure}[h]
\begin{center}
\includegraphics[width=16cm]{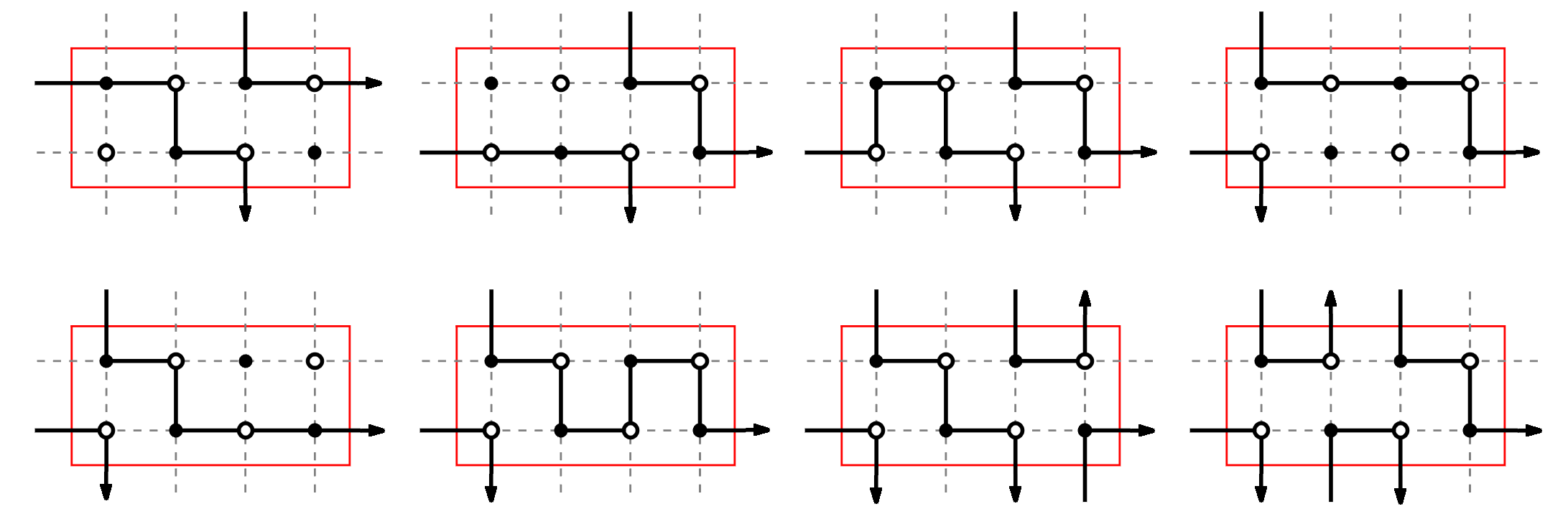}
\caption{\label{fig:empty1} Closing constraint: the sum of these loop variables must vanish.
}
\end{center}
\end{figure}

\begin{figure}[h]
\begin{center}
\includegraphics[width=16cm]{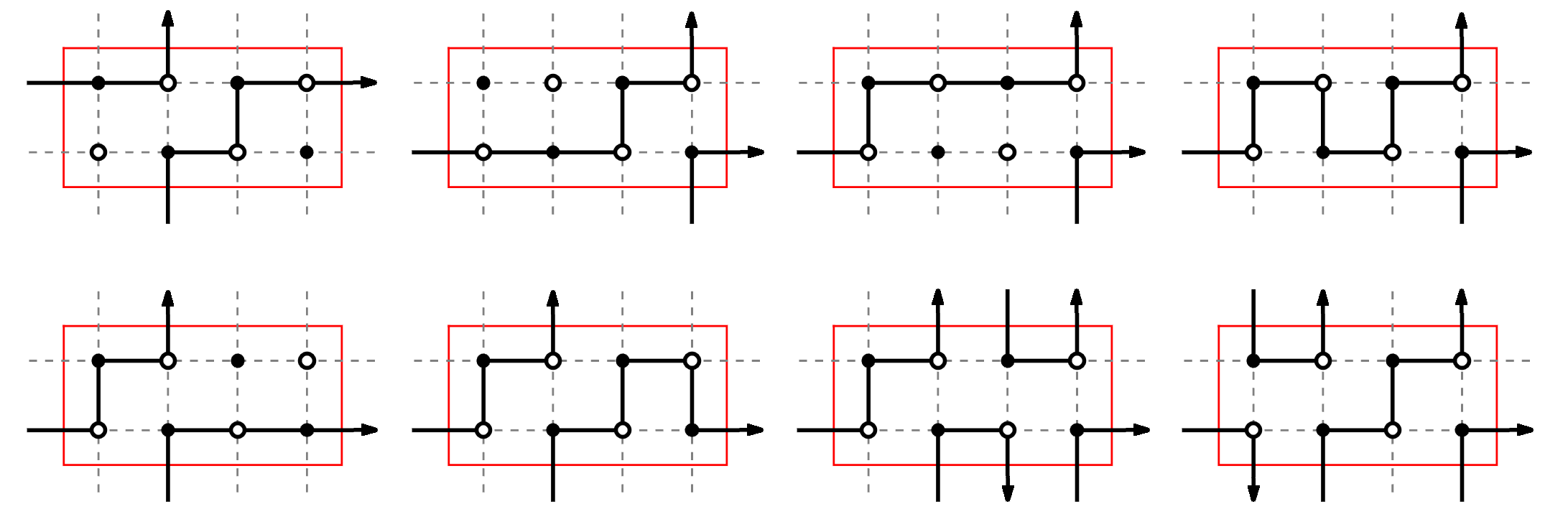}
\caption{\label{fig:empty2} Closing constraint: the sum of these loop variables must vanish.
}
\end{center}
\end{figure}

The resulting set of equations seems quite complicated in general and is left for further investigation. For $n=4$ segments a simple constraint is shown in Figure \ref{fig:closingeq1}. This arises from setting the difference of the linear terms in (\ref{eq:lin1}) and (\ref{eq:lin2}) to zero. Other constraints are shown in Figures \ref{fig:empty1}, \ref{fig:empty2}. These equations are responsible for the vanishing of the Hamiltonians indicated by empty circles in Figure \ref{fig:toric4}.

\clearpage

\clearpage

\begin{figure}[h]
\begin{center}
\includegraphics[width=4cm]{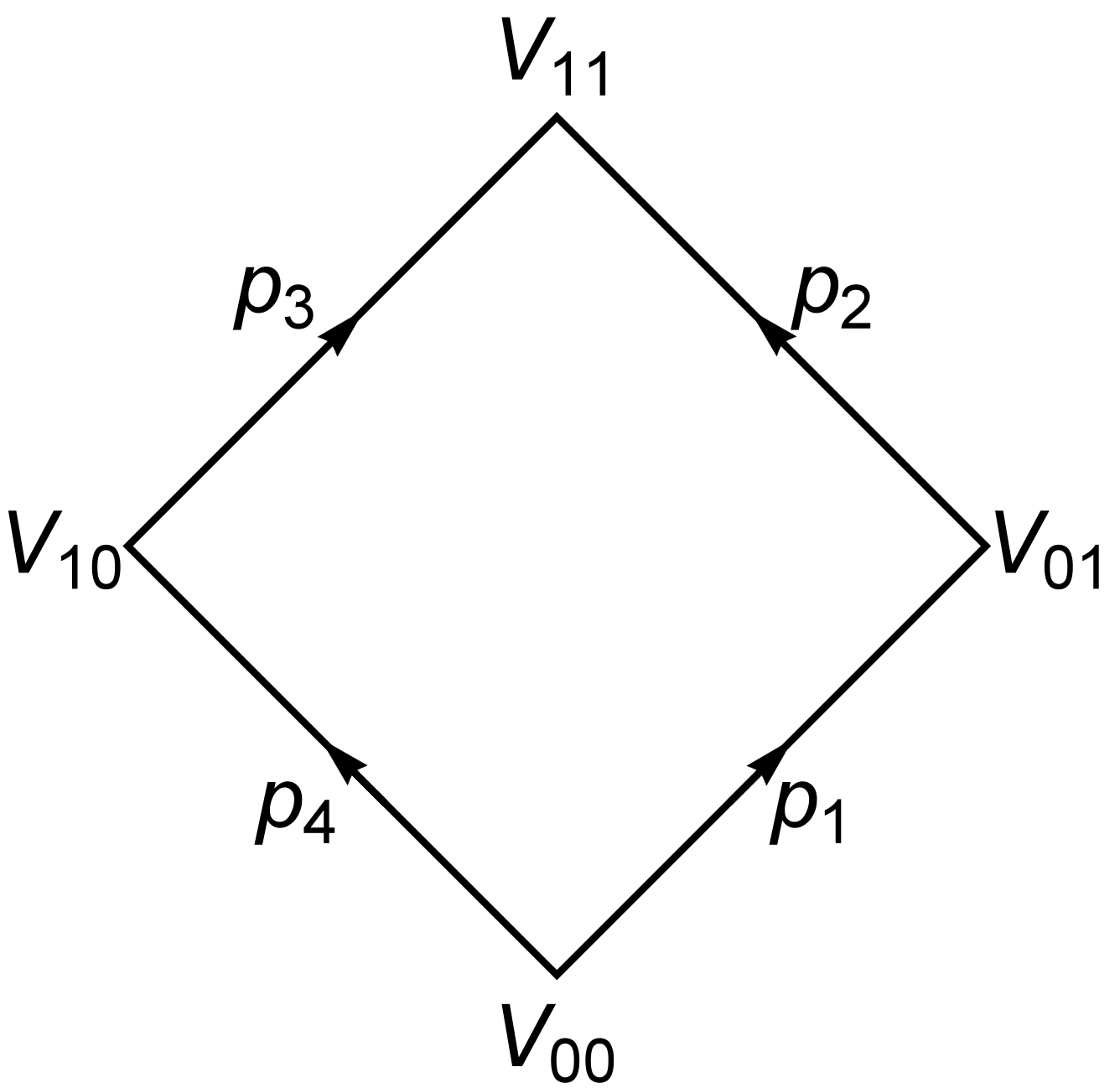}
\caption{\label{fig:momcon} An elementary patch of the worldsheet. The patch is bounded by four kink  worldlines. The tangent vectors of the worldlines are constant vectors $p_i \in \RR^{2,2}$ and $p_i^2 = 0$.
}
\end{center}
\end{figure}

\section{Coupling to a background B-field}

The worldsheet can be coupled to a background B-field whose field strength is the volume form. In this section we discuss the consequences of this coupling. The discrete equation (\ref{eq:deqn}) does not change \cite{Vegh:2016fcm}, but turning on the coupling deforms the constraints responsible for the closure of the string.

Our starting point is the action
\be
  \label{eq:action}
  S  = -{T \ov 2}\int d^2 \sigma \, \p_a Y^M \p_b Y^N (\sqrt{-h}h^{ab} G_{MN} + \kappa \epsilon^{ab} B_{MN}) \, .
\ee
Here $Y^M$ are coordinates on AdS$_3$, $h$ and $G$ are the worldsheet and background metrics, respectively. $B$ is the background two-form with a field strength that is proportional to the volume form of AdS$_3$.
Finally, $\kappa$ is its coupling to the worldsheet. It can be shown \cite{Gubser:2016zyw}, that causality demands $\kappa \in [-1,1]$. Without loss of generality we can take $\kappa \ge 0$.

Let us consider a worldsheet patch as in FIG. \ref{fig:momcon}. The four vertices  are labeled by $V_{ij} \in \RR^{2,2}$ where $ V_{ij}^2 = -1$.
Let us define the following kink tangent vectors
\bea
  \nonumber
   p_1 =  V_{01} -  V_{00} &\qquad &
   p_2 =  V_{11} -  V_{01} \\
  \label{eq:difvec}
   p_3 =  V_{11} -  V_{10} &\qquad &
   p_4 =  V_{10} -  V_{00}
\eea

Let $X(\sigma^-, \sigma^+) \in \RR^{2,2} $ denote the embedding function of the string into spacetime where $\sigma^\pm$ are lightcone coordinates on the worldsheet.
The patch is bounded by
\be
  \nonumber
  X(\sigma^-, 0) = V_{00} + \sigma^- p_4  \, ,  \qquad
  X(0, \sigma^+) = V_{00} + \sigma^+ p_1 \, .
\ee
for $\sigma^\pm \in (0,1)$.
Points inside the patch are given by the {\it interpolation ansatz} \cite{Gubser:2016zyw} which solves the equations of motion
\bea
  \nonumber
  X (\sigma^-, \sigma^+) =  {1+ (1+\kappa^2)\sigma^- \sigma^+ p_4 \cdot p_1 /2 \over
  1 - (1-\kappa^2)\sigma^- \sigma^+ p_4 \cdot p_1 /2} V_{00}
    + {\sigma^-p_4 + \sigma^+ p_1 + \kappa \sigma^+\sigma^-  N \over 1-(1-\kappa^2)\sigma^+\sigma^-  p_4 \cdot p_1 /2}
  \label{eq:ian}
\eea
where  
\be
  \nonumber
  N^\mu \equiv \epsilon^\mu_{\nu\lambda\rho} V^\nu_{00} p_4^\lambda p_1^\rho
\ee
and $\epsilon$ is the Levi-Civita tensor. For $\kappa=0$, the interpolation ansatz gives an AdS$_2$ patch embedded into AdS$_3$ and $N$ is proportional to the unit normal vector. For $\kappa \ne 0$, the normal vector is not constant and $N$ is only perpendicular to the surface at $V_{00}$.
From the interpolation ansatz, we have
\be
  V_{11} = X(1,1) \, .
\ee
The set of points in AdS$_3$ that are null separated from both $V_{10}$ and $V_{01}$ form a one-dimensional submanifold, conveniently parametrized by $V_{11}(\kappa)$.

\subsection{Modified normal vectors}

Although $N$ fails to be a normal vector for the patch when $\kappa \ne 0$, it can be shown that there exists a vector $\tilde N$ that satisfies
\be
  \label{eq:ferde}
  \tilde N X (\sigma^-, \sigma^+) = {\kappa \ov \sqrt{1-\kappa^2}} \, .
\ee
The argument goes as follows. Let us pick a generic $V_{00}$ point,
\be
  \nonumber
  V_{00} = (q_{-1}, \, q_{0}, \, q_{1}, \, q_{2} ) \, ,
\ee
with $V_{00}^2 = -1$, and two celestial fields, $b$ and $w$, corresponding to the vectors $p_4$ and $p_1$, respectively. These data fix the direction of both $p_4$ and $p_1$,
\be
  p_4 = {A \ov b-w}
  \begin{pmatrix}
-(1+b^2)q_0+(1-b^2)q_1+2 b q_2 \\
(1+b^2)q_{-1}-2b q_1 + (1-b^2)q_2 \\
(1-b^2)q_{-1} - 2b q_0 + (1+b^2)q_2 \\
2 b q_{-1} + (1-b^2)q_0 - (1+b^2)q_1
\end{pmatrix} \, ,
\ee
\be
  p_1 = {B \ov b-w}
  \begin{pmatrix}
-(1+w^2)q_0+(1-w^2)q_1+2 w q_2 \\
(1+w^2)q_{-1}-2w q_1 + (1-w^2)q_2 \\
(1-w^2)q_{-1} - 2w q_0 + (1+w^2)q_2 \\
2 w q_{-1} + (1-w^2)q_0 - (1+w^2)q_1
\end{pmatrix} \, ,
\ee
where $A,B$ are two constants. Using the interpolation ansatz, we now define $\tilde N$ to be a unit vector satisfying (\ref{eq:ferde}) at three points,
\be
   \tilde N V_{00} =   \tilde N X (0, 1) =   \tilde N X (1, 0) = {\kappa \ov \sqrt{1-\kappa^2}} \, .
\ee
We get
\be
  \tilde N = {1 \ov (b-w) \sqrt{1-\kappa^2}}
  \begin{pmatrix}
 \kappa(w-b)q_{-1}+  (1+b w)q_0 + (bw-1)q_1 - (b+w)q_2 \\
 -(1+bw)q_{-1} + \kappa (w-b) q_0 + (b+w)q_1 + (bw-1)q_2 \\
 (bw-1)q_{-1} + (b+w)q_0 + \kappa(w-b)q_1 - (1+bw)q_2 \\
 -(b+w)q_{-1} + (bw-1)q_0 + (1+bw)q_1 + \kappa(w-b)q_2
\end{pmatrix} \, .
\ee
A straightforward calculation shows that $\tilde N$ satisfies (\ref{eq:ferde}) on the entire patch.

\subsection{Modified reflection matrices}

The reflection matrix from section \ref{sec:recons} can be modified to include the contribution from the B-field if instead of (\ref{eq:nemferde}) we demand
\be
  \mathcal{R}_{b,w} V_{00} = \tilde N \, .
\ee
This gives
\be
\label{eq:reflmatx}
\mathcal{R}_{b,w}(\kappa) =
{1\ov (b-w)\sqrt{1-\kappa^2}}
\begin{pmatrix}
(w-b)\kappa & bw+1 & bw-1 & -b-w
\cr
-1-bw & (w-b)\kappa & b+w & bw-1
\cr
-1+bw & b+w & (w-b)\kappa & -1-bw
\cr
-b-w & bw-1 & bw+1 & (w-b)\kappa
\end{pmatrix}   \, .
\ee
The matrix satisfies
\be
  \nonumber
  \mathcal{R}_{b,w}(-\kappa) = \mathcal{R}^{-1}_{b,w}(\kappa) \, .
\ee
The same exercise can be done using the right-handed celestial fields $\tilde b, \tilde w$. The result is another matrix $\mathcal{\widetilde R}$  satisfying
\be
  \mathcal{\widetilde R}_{\tilde b, \tilde w} V_{00} = \tilde N \, .
\ee
A calculation similar to the previous one gives
\be
\label{eq:reflmaty}
\mathcal{\widetilde R}_{\tilde b,\tilde w}(\kappa) =
{1\ov (\tilde b-\tilde w)\sqrt{1-\kappa^2}}
\begin{pmatrix}
(\tilde w - \tilde b)\kappa & -1-\tilde b\tilde w & 1-\tilde b\tilde w & -\tilde b-\tilde w
\cr
1+\tilde b\tilde w & (\tilde w - \tilde b)\kappa & -\tilde b-\tilde w & \tilde b\tilde w-1
\cr
1-\tilde b\tilde w & -\tilde b-\tilde w & (\tilde w - \tilde b)\kappa & -1-\tilde b\tilde w
\cr
-\tilde b-\tilde w & \tilde b\tilde w-1 & \tilde b\tilde w+1 & (\tilde w - \tilde b)\kappa
\end{pmatrix}   \, .
\ee
The matrix satisfies
\be
  \nonumber
  \mathcal{\widetilde R}_{\tilde b,\tilde w}(-\kappa) = \mathcal{\widetilde R}^{-1}_{\tilde b,\tilde w}(\kappa) \, .
\ee

At $\kappa=\pm 1$ both (\ref{eq:reflmatx}) and (\ref{eq:reflmaty}) are singular. In this case one can only reconstruct the embedding  if both left- and right-handed celestial fields are known. This case corresponds to the $SL(2)$ WZW model.

\subsection{Closing constraints}

Since the coupling to the B-field changes $\mathcal{ R}$, the closing constraints also change. In the following we will study the constraints at the level of the spectral curve. Let us consider a closed string in Figure \ref{fig:spectral} (right) and we now assume that the string couples to the B-field. From $Q$ we compute $N_2$ and so on until we get to $P'$. The final result is $P' = \mathcal{R} P$ where
\be
  \label{eq:product}
   \mathcal{R} =   \mathcal{R}_{b_N,w_N} \cdots \mathcal{R}_{b_2,w_2} \mathcal{R}_{b_1,w_1} \, .
\ee
If $\mathcal{R}  = \textrm{Id}_{4\times 4}$ then we have $P=P'$, i.e. the string closes in target space.
Similarly to the analysis in \cite{Vegh:2021jhl} we now rewrite this constraint using a product of $\Omega_{b,w}$  matrices as follows.
The $n^\textrm{th}$ matrix in the product can be decomposed as
\be
   \nonumber
  \mathcal{R}_{b_n,w_n} =  {c^\mu_n \Sigma_\mu  }\, ,
\ee
where summation over $\mu \in \{ 0, 1,2 , 3\}$ is understood. The constants are given by%
\be
  \nonumber
  c^\mu_n = {1\ov  \sqrt{1-\kappa^2}}\le( (-1)^n \kappa  , \
  {1- b_n w_n \ov b_n-w_n}, \
   i{1+b_n w_n \ov w_n - b_n},  \
   {b_n + w_n \ov w_n - b_n }  \ri)\, ,
\ee
and the $\Sigma_\mu$ matrices are defined by
\be
\nonumber
  \Sigma_0 = \mathbb{1}_{4\times 4}
\, , \quad
\Sigma_1 =
\begin{pmatrix}
0 & 0 & \hskip -0.2cm -1 & 0
\cr
0 & 0 & 0 & \hskip -0.2cm \hskip -0.0cm -1
\cr
-1 & 0 & 0 & 0
\cr
0 & \hskip -0.2cm -1 & 0 & 0
\end{pmatrix}   \, , \quad
\Sigma_2 =
\begin{pmatrix}
0 & i & 0 & 0
\cr
-i & 0 & 0 & 0
\cr
0 & 0 & 0 & -i
\cr
0 & 0 & i & 0
\end{pmatrix}   \, , \quad
\Sigma_3 =
\begin{pmatrix}
0 & 0 & 0 & 1
\cr
0 & 0 & \hskip -0.2cm -1 & 0
\cr
0 & \hskip -0.2cm -1 & 0 & 0
\cr
1 & 0 & 0 & 0
\end{pmatrix}   \, .
\ee
These matrices satisfy $\Sigma_i \Sigma_j = \delta_{ij} \Sigma_0 + i \varepsilon_{ijk} \Sigma_k$,
and thus they generate an algebra isomorphic to the one generated by $\sigma_0 =\mathbb{1}$ and the three Pauli matrices.
We can thus replace the $\Sigma$ matrices in $\mathcal{R}$ with their $2 \times 2$ counterparts and define
\be
  \nonumber
  \mathcal{R}'  = \prod_{n=N}^1   c^\mu_n \sigma_\mu  \, .
\ee
We have $\mathcal{R}'=\mathbb{1}$ precisely when $\mathcal{R} = \Sigma_0$.
For odd $n$ we can write
\be
  \nonumber
    c^\mu_n \sigma_\mu =   i \sigma_3 \Omega_{b_n,w_n}(\zeta=\zeta_0)\sigma_3 \, ,
\ee
and for even $n$ we write
\be
  \nonumber
    c^\mu_n \sigma_\mu =     -i \sigma_3\Omega^{-1}_{b_n,w_n}(\zeta=\zeta_0)\sigma_3 \, ,
\ee
where we defined the constant
\be
  \nonumber
  \zeta_0 =i {1+\kappa \ov \sqrt{1-\kappa^2}} \, .
\ee
The $\sigma_3$ matrices and the constant $\pm i$ factors cancel in $ \mathcal{R}'$, which then gives the Lax monodromy from $P$ to $P'$ (sandwiched between two $\sigma_3$ matrices).
Thus, the string closes precisely when the monodromy at $\zeta = \zeta_0$  is trivial, i.e.
\be
  \nonumber
  \Omega(\zeta = \zeta_0) =   \mathbb{1} \, .
\ee
Since the equation of motion remains the same, this is the only effect that the coupling to the two-form has on the string in our description.

\clearpage

\begin{figure}[h]
\begin{center}
\includegraphics[width=6.6cm]{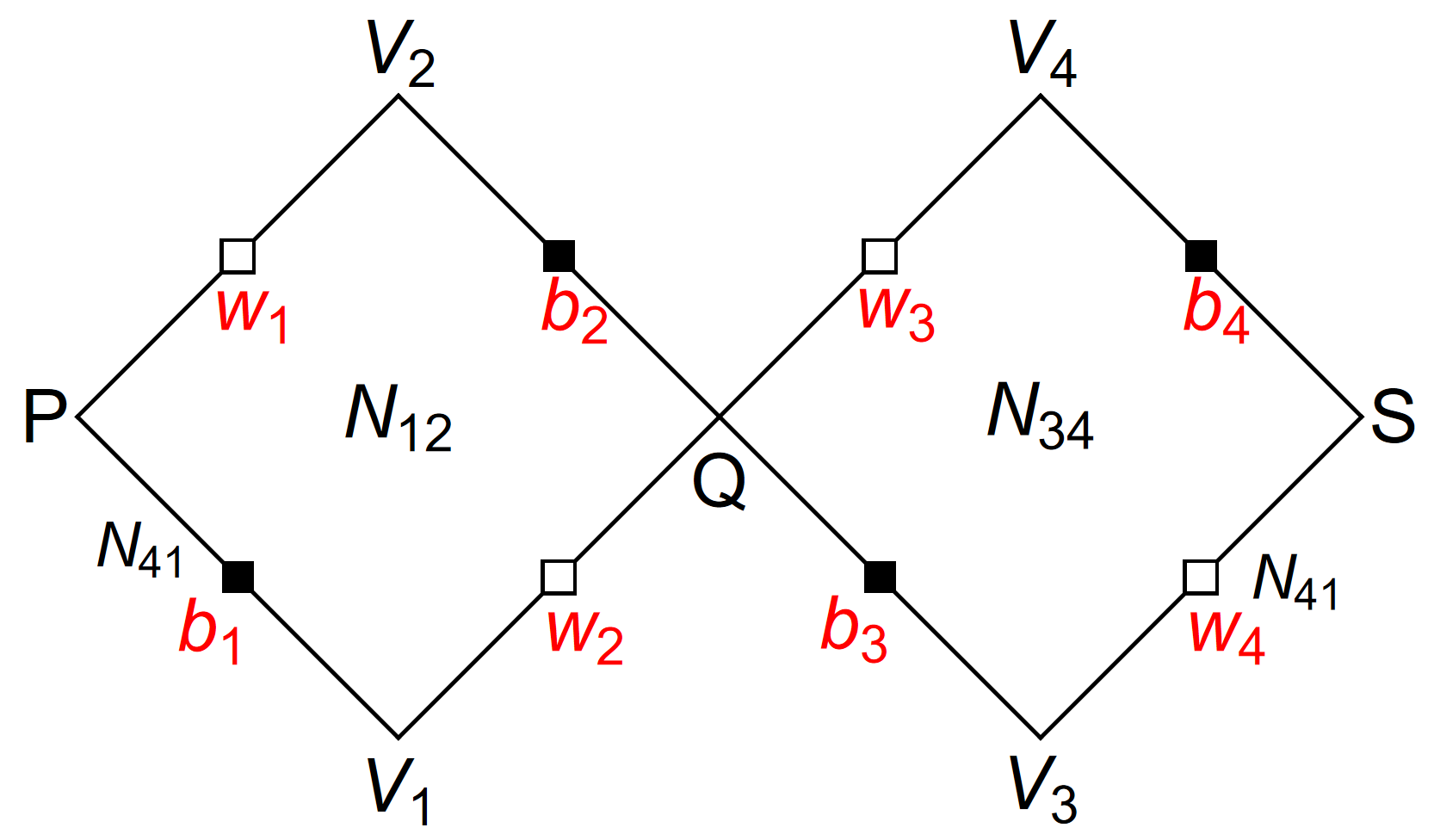}
\caption{\label{fig:tildevars} Constructing face loop variables from right-handed celestial variables. $N_{12}, N_{23}$ are normal vectors, $b_i$ and $w_j$ are left-handed celestial variables. $P, Q, S, V_i$ are spacetime points.
}
\end{center}
\end{figure}

\subsection{Right-handed celestial fields}

In this last part, let us discuss the relationship of the right-handed celestial variables to the $x_i,y_i, p_i, q_i$ variables. The latter can be computed from cross-ratios of nearby celestial variables. We will need two AdS$_2$ patches as shown in Figure \ref{fig:tildevars}. Let us assume that the left-handed celestial variables are known and that $P$ is a generic point in $\RR^{2,2}$ and $P^2 = -1$. Using $\mathcal{R}_{b,w}(\kappa)$ we calculate all the other spacetime points and normal vectors,
\bea
  \nonumber
  N_{12} =  \mathcal{R}_{b_1,w_1}(\kappa) P \, , \qquad
  & V_{1} =  \mathcal{R}_{b_1,w_2}(-\kappa) N_{12} \, , & \qquad
  V_{2} =  \mathcal{R}_{b_2,w_1}(-\kappa) N_{12}  \, , \\
  Q  =  \mathcal{R}_{b_2,w_2}(-\kappa) N_{12} \, , \qquad
  & N_{34} =  \mathcal{R}_{b_3,w_3}(\kappa) Q \, , & \qquad
  V_{3} =  \mathcal{R}_{b_3,w_4}(-\kappa) N_{34}  \, , \\
  V_{4} =  \mathcal{R}_{b_4,w_3}(-\kappa) N_{34}\, , \qquad
  & S =  \mathcal{R}_{b_4,w_4}(-\kappa) N_{34} \, , &
\eea
and from (\ref{eq:aeq}) the right-handed celestial variables can be computed by taking differences of spacetimes points, e.g.
\be
  \nonumber
  \tilde b_1 = {X_{-1} + X_2 \over -X_0 + X_1}  \, , \qquad \textrm{where} \ X \equiv V_1 - P \, .
\ee
The $x_i, y_i, p_i, q_i$ variables can be computed from (\ref{eq:xdef}), (\ref{eq:ydef}), (\ref{eq:pdef}), (\ref{eq:qdef}), respectively. If we instead plug in  the right-handed variables, then we obtain the quantities that we denote $\tilde x_i, \tilde y_i, \tilde p_i, \tilde q_i$.
After a lengthy but straightforward calculation we get
\be
  \nonumber
   \tilde p_i= p_i \, , \qquad   \tilde q_i = q_i \, , \qquad
   \tilde x_i = {\kappa-1 \ov \kappa+1} x_i \, , \qquad
   \tilde y_i = {\kappa-1 \ov \kappa+1} y_i \, .
\ee
Even though $x_i,y_i$ and $\tilde x_i, \tilde y_i$ differ by constant factors\footnote{Note that the equations are symmetric under a $\ZZ_2$ parity transformation that exchanges $a \leftrightarrow \tilde a$ and flips the sign of $\kappa$.},  the same $p_i,q_i$ dynamical variables are associated to the string independently of whether we compute them from left- or right-handed celestial fields.

\clearpage

\section{Discussion}

This paper described the embedding of segmented strings in AdS$_3$ into the cluster transformation dynamics of $Y^{n,0}$  brane tilings (a.k.a. dimer models).
Brane tilings are doubly periodic planar bipartite graphs, which appeared in the study of four-dimensional quiver gauge theories on the worldvolume of D3-branes probing non-compact toric Calabi-Yau threefolds \cite{Hanany:2005ve, Franco:2005rj, Hanany:2005ss, Feng:2005gw}. Goncharov and Kenyon showed \cite{Goncharov2013} that every (consistent) brane tiling defines an integrable system, whose constants of motion can be systematically calculated from perfect matchings.
Dynamical variables are loops in the tiling, which in the case of $Y^{n,0}$ can been computed from the string's $\RR^{2,2}$ celestial embedding variables. These variables are constructed from the null velocity vectors of the kinks that propagate on the segmented string.
The tiling technology allowed for an alternative calculation of the spectral curve in terms of celestial variables. The results have been matched with previous calculations, which used matched asymptotic expansions \cite{Vegh:2021jhl}.

Closed strings have periodic embedding functions as $\sigma \to \sigma + 2\pi$ and likewise periodic celestial fields. Loop variables of the tiling are computed by taking certain cross-ratios of the celestial fields and thus they are invariant under the $SO(2,2)$ AdS$_3$ isometry group. If they take generic values, then the corresponding celestial fields are non-periodic and thus the string embedding has a monodromy in target space: it only closes up to an $SO(2,2)$ transformation. Demanding a trivial monodromy imposes constraints on the loop variables. Some of the constraints have been identified in terms of these variables, but a complete study is left for future work.
Finally, we have discussed how to deform the constraints by turning on a coupling to a background two-form whose field strength is proportional to the volume form.

The description of string embeddings used celestial fields, which are computed from vector components in the ambient space $\RR^{2,2}$ as discussed in section \ref{sec:cele}. Since AdS$_3$ is defined as the universal cover of a hyperboloid in this space, there is the potential danger of strings wrapping closed timelike cycles.
Although we have not discussed it in this paper, eliminating such embeddings imposes further constraints on the variables.

An immediate question for the future is how to quantize the segmented string. The Goncharov-Kenyon quantum integrable model corresponding to $Y^{n,0}$  is the relativistic Toda model. In order to quantize the physical string, one needs to take into account the closing constraints. Presumably this can be done at least in certain cases. It is expected that recent developments on exact quantization will be of use (see e.g. \cite{Franco:2015rnr} for conjectures on exact quantization conditions in the context of brane tilings).

Loop variables corresponding to faces in the brane tiling undergo cluster transformations at each discrete time step. The equations governing these variables form a Y-system \cite{ZAMOLODCHIKOV1991391}. Y-systems   have previously appeared  in the context of AdS/CFT \cite{Gromov:2009tv, Bombardelli:2009ns, Gromov:2009bc, Arutyunov:2009ur, Gromov:2009tq, Gromov:2011cx} where they have been used to  compute the planar spectrum of anomalous dimensions in $\mathcal{N}=4$ SYM theory (for a review see \cite{Gromov:2010kf, Bajnok:2010ke}).
Planar scattering amplitudes at strong coupling correspond to minimal surfaces that end on a null polygonal contour at the AdS boundary \cite{Alday:2007hr}. Y-systems have also been used in this context \cite{Alday:2010vh}. Note that in these applications the boundary conditions for the Y-functions are fixed (non-periodic). It would be interesting to relate these results to segmented strings.

Several interesting generalizations may be possible. One could consider a segmented string in BTZ \cite{Gubser:2016wno}, AdS$_n$, or dS$_n$  geometries and, if exists, find the corresponding brane tiling. The $Y^{n,0}$ quiver is the `product' of Dynkin diagrams of the $\hat A_1$ and $\hat A_n$ affine Lie algebras. Strings in AdS$_n$ are governed by other affine Toda models whose Dynkin diagrams might give a hint for what the relevant brane tiling might be.

If one does not insist on the interpretation of a string moving in a given target space, then one can replace $Y^{n,0}$  with generic brane tilings or with products of generic (affine) Lie algebras \cite{Yamazaki:2016aam} (see also \cite{Turiaci:2016cvo}).
It would be interesting to give a meaning to the `duality' whereby one rotates the tiling by 90$^\circ$. Perhaps this transformation can be interpreted as a holographic duality since the model loses its discrete $\sigma$ direction.

\vspace{0.2in}   \centerline{\bf{Acknowledgments}} \vspace{0.2in}
I am grateful to Andrea Cavagli\` a for a useful discussion,  
and I thank  the organizers of the
Physics Sessions Initiative in Crete for providing a stimulating environment.
The author is supported by the STFC Ernest Rutherford grant ST/P004334/1.

\clearpage

\appendix

\begin{figure}[h]
\begin{center}
\includegraphics[width=3.5cm]{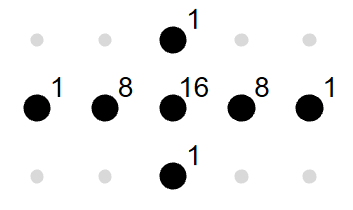}
\caption{\label{fig:y40nums} Newton polygon with perfect matching multiplicities for $Y^{4,0}$.
}
\end{center}
\end{figure}

\section*{Perfect matchings for $Y^{4,0}$}

\begin{figure}[h]
\begin{center}
\vskip-0.5cm  \includegraphics[width=16cm]{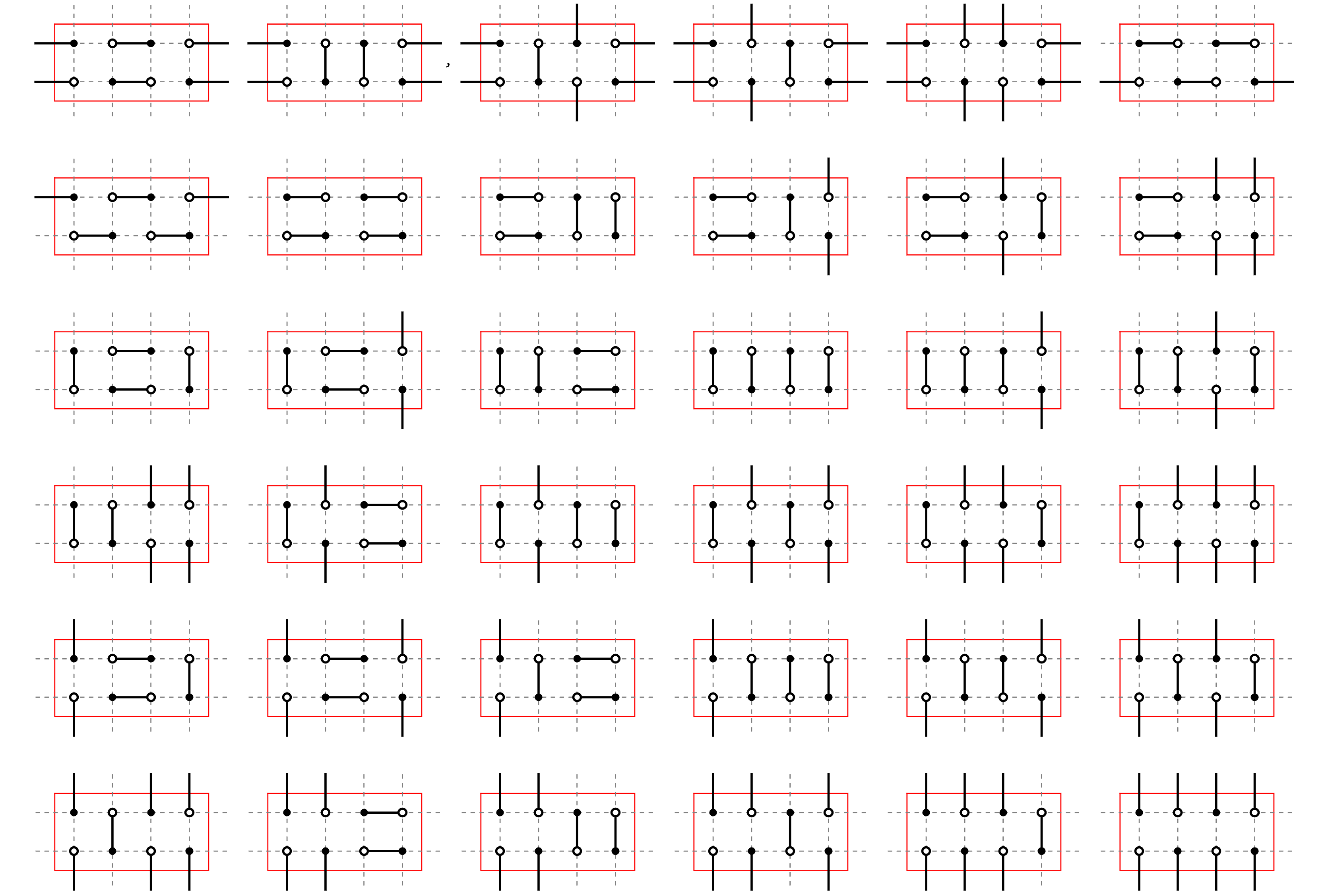}
\caption{\label{fig:y40pms} All 36 perfect matchings for $Y^{4,0}$. They correspond to points in the Newton polygon in Figure \ref{fig:y40nums}.
}
\end{center}
\end{figure}

The determinant of the Kasteleyn matrix without the edge weights counts the perfect matchings. In the general $n$ case it is given by \cite{Franco:2005rj}
\be
\nonumber
K = \left( \begin{array}{cccccccc}
x-1 & 1 & 0 & 0 & \hdotsfor{3} & y \\
1 & 1-x^{-1} & 1 & 0 & \hdotsfor{3} & 0\\
0 & 1 & x-1 & 1 & 0 & \hdotsfor{2} & 0\\
0 & 0 & 1 & 1-x^{-1} & 1 & 0 & \ldots  & 0\\
& \vdots & & & & \ddots & & \\
0 & \hdotsfor{3} & 0 & 1 & x-1 & 1 \\
y^{-1} & 0 & \hdotsfor{3} & 0 & 1 & 1-x^{-1}
\end{array} \right)
\ee
For $n=4$, the Newton polygon of the determinant is shown in Figure \ref{fig:y40nums}. The diagram shows the (non-zero) coefficients corresponding to each lattice point. These correspond to perfect matching multiplicities. Altogether there are 36 perfect matchings listed in Figure~\ref{fig:y40pms}.

\clearpage
\bibliographystyle{JHEP}
\bibliography{paper}

\end{document}